\documentclass[preprint,showpacs,preprintnumbers,amsmath,amssymb]{revtex4}

\usepackage{graphicx}
\usepackage{dcolumn}
\usepackage{bm}
\usepackage{float,color}

\newcommand{\bff}[1]{{\mbox{\boldmath $#1$}}}

\newcommand{\bra}[1]{\left\langle #1 \right|}
\newcommand{\ket}[1]{\left| #1 \right\rangle}

\begin{document}

\title{Implementation of the finite amplitude method for the relativistic quasiparticle random-phase approximation}

\author{T. Nik\v{s}i\'{c}$^1$, N. Kralj$^{1}$, T. Tuti\v{s}$^{1}$, D. Vretenar$^1$, P. Ring$^2$}
\affiliation{$^{1}$Physics Department, Faculty of Science, University of Zagreb, 10000 Zagreb, Croatia}
\affiliation{$^{2}$Physik-Department der Technischen Universit\"at M\"unchen, D-85748 Garching, Germany}

\date{\today}

\begin{abstract} 
A new implementation of the finite amplitude method (FAM) for the solution of the 
relativistic quasiparticle random-phase approximation (RQRPA) is presented, 
based on the relativistic Hartree-Bogoliubov (RHB) model for deformed 
nuclei. The numerical accuracy and stability of the FAM -- RQRPA 
is tested in a calculation of the monopole response of $^{22}$O. 
As an illustrative example, the model is applied to a study of 
the evolution of  monopole strength in the chain of Sm isotopes, 
including the splitting of the giant monopole resonance in axially deformed systems.
\end{abstract}


\maketitle

\section{\label{sec-introduction}Introduction}

Vibrational modes and, more generally, collective degrees of freedom have been a 
recurring theme in nuclear structure studies over many decades. An especially 
interesting topic that has recently attracted considerable interest is the multipole 
response of nuclei far from stability and the possible occurrence of exotic 
modes of excitation \cite{Nils.07,SAZ.13}. For theoretical studies of collective 
vibrations in medium-heavy and heavy nuclei the tool of choice is the 
random-phase approximation (RPA), or quasiparticle random-phase 
approximation (QRPA) for open-shell nuclei \cite{RS.80}. The (Q)RPA 
equations can be obtained by linearizing the time-dependent 
Hartree-Fock (Bogoliubov) equations, and the standard method of solution 
presents a generalized eigenvalue problem for the (Q)RPA matrix. 
As the space of quasiparticle excitations can become very large 
in open-shell heavy nuclei, the standard matrix solution of the 
QRPA equations is often computationally prohibitive, especially 
for deformed nuclei. In addition to the fact that a very large number 
of matrix elements has to be computed, the calculation of each 
matrix elements is complicated by the fact that most modern 
implementations of the (Q)RPA are fully self-consistent, that is, 
the residual interaction is obtained as a second functional derivative 
of the nuclear energy density functional with respect to the nucleonic 
one-body density. Since energy density functionals can be quite 
complicated and include many terms \cite{BHR.03,EDF.04}, this 
produces complex residual interactions and 
the computation of huge (Q)RPA matrices becomes excessively 
time-consuming.

Although several new implementations of the fully self-consistent matrix QRPA 
for axially deformed nuclei have been developed in recent 
years \cite{PG.08,YG.08,Pena.09,TE.10,Losa.10}, 
and even applied to studies of collective modes in rather heavy deformed nuclei, 
the huge computational cost has so far prevented systematic studies of 
multipole response in deformed nuclei.  An interesting and very useful 
alternative solution of the (Q)RPA problem has recently been proposed, based 
on the finite-amplitude method (FAM) \cite{NIY.07} . In this approach one 
avoids the computation and diagonalization of the (Q)RPA matrix by 
calculating, instead, the fields induced by the external one-body operator 
and iteratively solving the corresponding linear response problem. 
The FAM for the RPA has very successfully been employed in a self-consistent 
calculation of nuclear photo absorption
cross sections \cite{INY.09}, and in a study of the emergence of 
pygmy dipole resonances in nuclei far from stability \cite{INY.11}.
More recently the FAM has been extended to the quasiparticle 
RPA based on the Skyrme Hartree-Fock-Bogoliubov (HFB) 
framework \cite{AN.11,Sto.11,HKN.13}. The feasibility of the finite amplitude 
method for the relativistic RPA has been investigated in Ref.~\cite{Liang.13}.

In this work we report a new implementation of the FAM for the 
relativistic quasiparticle random-phase approximation (RQRPA), based 
on the relativistic Hartree-Bogoliubov (RHB) model for mean-field studies 
of deformed open-shell nuclei  \cite{VALR.05,Meng.06}. The standard matrix 
RQRPA for spherical nuclei was formulated in the canonical single-nucleon basis
of the RHB model \cite{Paar2003}, extended to the description of
charge-exchange excitations (pn-RQRPA) in Ref.~\cite{Paar2004}, 
and further extended to deformed systems with axial symmetry in 
Ref.~\cite{Pena.09}. Here we develop a FAM method for the 
small-amplitude limit of the time-dependent Hartree-Bogoliubov  
framework based on relativistic energy density functionals 
and a pairing force separable in momentum space, and perform 
tests and illustrative calculations for the new model. 
 
The paper is organized as follows. In Sec.~\ref{TDRHB-QRPA} we briefly recapitulate 
the small-amplitude limit of the time-dependent RHB model and present a new 
implementation of the FAM for this particular framework. 
Numerical details and test calculations are included in Sec.~\ref{sec-numerical},
and in Sec.~\ref{sec-illustrative-calculations} we apply the model to a study of 
the evolution of  monopole strength in the chain of Sm isotopes. Section~\ref{sec-summary}
summarizes the results and ends with an outlook for future applications. Details
on the expansion of single-nucleon spinors in the axially symmetric harmonic oscillator
basis, calculation of the matrix elements monopole operator, time-odd terms in the FAM equations,
and RHB and FAM equations with time-reversal symmetry, are included in 
Appendix~\ref{Sec-App1}-\ref{Sec-App4}. 
\section{\label{TDRHB-QRPA}Small amplitude limit of the time-dependent RHB model 
and the finite amplitude method}

The relativistic Hartee-Bogoliubov (RHB) model \cite{VALR.05,Meng.06}
provides a unified description of nuclear particle-hole $(ph)$ and particle-particle
$(pp)$ correlations on a mean-field level by combining two average
potentials: the self-consistent nuclear mean field that
encloses all the long range \textit{ph} correlations, and a
pairing field $\hat{\Delta}$ which sums up the
\textit{pp}-correlations.
In the RHB framework the nuclear single-reference state is described
by a generalized Slater determinant
$|\Phi\rangle$ that represents a vacuum with respect to independent
quasiparticles. The quasiparticle operators are defined by the
unitary Bogoliubov transformation,  and the corresponding  Hartree-Bogoliubov wave
functions $U$ and $V$ are determined by the solution of the RHB equation:
\begin{equation}
\label{eq:RHB}
\left( \begin{array}{cc}
   h_D-m-\lambda  &  \Delta \\
   -\Delta^*  & -h_D^*+m+\lambda
\end{array} \right) \left( \begin{array}{c} U_k \\ V_k \end{array} \right)
= E_k    \left( \begin{array}{c} U_k \\ V_k \end{array} \right) \;.
\end{equation}
In the relativistic case the self-consistent mean-field is included in
the single-nucleon Dirac Hamiltonian $\hat{h}_D$, $\Delta$ is the pairing field,
and $U$ and $V$ denote Dirac spinors. In the formalism of  supermatrices
introduced by Valatin \cite{Valatin1961_PR122-1012}, the RHB functions are
determined by the Bogoliubov transformation which
relates the original basis of particle creation and
annihilation operators $c_{n}^{{}}$, $c_{n}^{\dag}$ (e.g. an oscillator basis) 
to the quasiparticle basis $\alpha_{\mu}^{{}}$, $\alpha_{\mu}^{\dag}$%
\begin{equation}
\left(
\begin{array}
[c]{c}%
c\\
c^{\dag}%
\end{array}
\right)  =\mathcal{W}\text{ }\left(
\begin{array}
[c]{c}%
\alpha\\
\alpha^{\dag}%
\end{array}
\right)  \text{\ \ \ \ \ \ \ with \ \ \ \ \ }\mathcal{W=}\left(
\begin{array}
[c]{cc}%
U & V^{\ast}\\
V & U^{\ast}%
\end{array}
\right)\;.
\end{equation}
In this notation a single-particle operator can be represented in the matrix form: 
\begin{equation}
\hat{F}=\frac{1}{2}\left(
\begin{array}
[c]{cc}%
\alpha^{\dag} & \alpha
\end{array}
\right)  \mathcal{F}\left(
\begin{array}
[c]{c}%
\alpha\\
\alpha^{\dag}%
\end{array}
\right)  + {\rm const.}\text{ \ \ \ \ }%
\end{equation}
with%
\begin{equation}
\mathcal{F=}\text{\ }\left(
\begin{array}
[c]{cc}%
F^{11} & F^{20}\\
F^{02} & {-(F^{11})}^\intercal%
\end{array}
\right) \;.
\end{equation}
In particular, for the generalized density $\mathcal{R}$:%

\begin{equation}
\mathcal{R}=\left(
\begin{array}
[c]{cc}%
\rho & \kappa\\
-\kappa^{\ast} & 1-\rho^{\ast}%
\end{array}
\right)\;,
\end{equation}
where the density matrix and pairing tensor read
\begin{equation}
\rho=V\mathcal{^{\ast}}V^{\intercal}\text{, \ \ \ \ \ \ }\kappa
=V\mathcal{^{\ast}}U^{\intercal},%
\end{equation}
respectively, and the RHB Hamiltonian is given by a functional derivative 
of a given energy density functional with respect to the generalized density:%
\begin{equation}
\mathcal{H=}\frac{\delta E[\mathcal{R}]}{\delta\mathcal{R}}=\left(
\begin{array}
[c]{cc}%
h & \Delta\\
-\Delta^{\ast} & -h^{\ast}%
\end{array}
\right)\;.
\end{equation}
The evolution of the nucleonic density subject to a time-dependent
external perturbation $\hat{F}(t)$ is determined by the 
time-dependent relativistic Hartree Bogolyubov (TDRHB) equation:
\begin{equation}
i\partial_{t}\mathcal{R}(t)\mathcal{=}\left[  \mathcal{H}(\mathcal{R}%
(t))+\mathcal{F}(t),\mathcal{R}(t)\right]\;.  \label{E7}%
\end{equation}
For a weak harmonic external field
\begin{equation}
\hat{F}(t)=\eta(\hat{F}(\omega)e^{-i\omega t}+\hat{F}^{\dag}(\omega)e^{i\omega t}),
\end{equation}
characterized by the small real parameter $\eta$, the density 
undergoes small-amplitude oscillations around the equilibrium  
with the same frequency $\omega$, that is, in the 
small-amplitude limit of the TDRHB:
\begin{equation}
\mathcal{R}(t)\mathcal{=R}_{0}+\eta(\delta\mathcal{R(\omega)}e^{-i\omega
t}+\delta\mathcal{R}^{\dag}\mathcal{(\omega)}e^{i\omega t})\;,
\end{equation}
and therefore
\begin{equation}
\mathcal{H}(t)\mathcal{=H}_{0}+\eta(\delta\mathcal{H(\omega)}e^{-i\omega
t}+\delta\mathcal{H}^{\dag}\mathcal{(\omega)}e^{i\omega t}) \;.
\end{equation}
The matrices $\delta\mathcal{R(\omega)}$ and
$\delta\mathcal{H(\omega)}$ are not necessarily Hermitian. By linearizing the
equation of motion (\ref{E7}) with respect to $\eta$, one obtains the linear-response 
equation in the frequency domain:
\begin{equation}
\omega~\mathcal{\delta\mathcal{R}=}\left[  \mathcal{H}_{0},\delta
\mathcal{R}\right]  +\left[  \delta\mathcal{H(\omega)},\mathcal{R}_{0}\right]
+\left[  \mathcal{F(\omega)},\mathcal{R}_{0}\right] \;.  \label{E11}%
\end{equation}
In the stationary quasiparticle basis the matrices $\mathcal{H}_{0}$ and $\mathcal{R}_{0}$ are
diagonal%
\begin{equation}
\mathcal{H}_{0}=\left(
\begin{array}
[c]{cc}%
E & 0\\
0 & -E
\end{array}
\right)  ,\text{ \ \ \ \ }\mathcal{R}_{0}=\left(
\begin{array}
[c]{cc}%
0 & 0\\
0 & 1
\end{array}
\right)\; ,
\end{equation}
and, because the density matrix is a projector ($\mathcal{R}^{2}\mathcal{=R}$) at all times, 
only the two-quasiparticle matrix elements of 
the time-dependent matrix $\delta\mathcal{R}$ do not vanish 
in this basis%
\begin{equation}
\delta\mathcal{R}=\left(
\begin{array}
[c]{cc}%
0 & R^{20}\\
R^{02} & 0
\end{array}
\right)  :=\left(
\begin{array}
[c]{cc}%
0 & X\\
Y & 0
\end{array}
\right)\;.  \label{E13}%
\end{equation}
This relation defines the QPRA amplitudes $X_{\mu\nu\text{ }}$ and $Y_{\mu\nu\text{ }}$.
In the quasiparticle basis Eq. (\ref{E11}) takes the form
\begin{eqnarray}
\mathcal{(}E_{\mu}+E_{\nu}-\omega)X_{\mu\nu}+\delta H_{\mu\nu}^{20} 
&=&-F_{\mu\nu}^{20}\\
\mathcal{(}E_{\mu}+E_{\nu}+\omega)Y_{\mu\nu}+\delta H_{\mu\nu}^{02} 
&=&-F_{\mu\nu}^{02} \;.%
\end{eqnarray}
Since $\delta\mathcal{H(\omega)}$ depends on $\delta\mathcal{R(\omega)}$, that is,
on the amplitudes $X_{\mu\nu\text{ }}$ and $Y_{\mu\nu\text{ }}$, this is actually a set
of non-linear equations. The expansion of $\delta H_{\mu\nu}^{20}$ and $\delta
H_{\mu\nu}^{02}$ in terms of $X_{\mu\nu\text{ }}$ and $Y_{\mu\nu\text{ }}$ up
to linear order leads to the conventional QRPA equations. These equations contain 
second derivatives of the density functional $E[\mathcal{R}]$ with respect to
$\mathcal{R}$ as matrix elements. For deformed nuclei in particular, the 
number of two-quasiparticle configurations
can become very large and the evaluation of matrix elements requires a considerable, 
and in many cases prohibitive, numerical effort. In many cases 
this has prevented systematic applications of the conventional 
QRPA method to studies of the multipole response of medium-heavy and 
heavy deformed nuclei. 

In the finite amplitude method for the QRPA \cite{AN.11,Sto.11}, the 
amplitude  $X_{\mu\nu\text{ }}$ and
$Y_{\mu\nu\text{ }}$ are formally expressed 
\begin{eqnarray}
X_{\mu\nu} &=&-\frac{F_{\mu\nu}^{20}+\delta H_{\mu\nu}^{20}}{E_{\mu}+E_{\nu}-\omega},\label{E16}\\%
Y_{\mu\nu} &=&-\frac{F_{\mu\nu}^{02}+\delta H_{\mu\nu}^{02}}{E_{\mu}+E_{\nu}+\omega},\label{E17}%
\end{eqnarray}
and $\delta\mathcal{H(\omega)}$ is calculated by numerical differentiation
\begin{equation}
\delta\mathcal{H(\omega)}=\lim_{\eta\rightarrow0}\frac{1}{\eta}(\mathcal{H}%
(\mathcal{R}_{0}+\eta\delta\mathcal{R(\omega)})-\mathcal{H}(\mathcal{R}_{0})) \;,
\label{E18}
\end{equation}
using a stationary RHB code for the evaluation of $\mathcal{H}(\mathcal{R})$.
We start from Eq.~(\ref{E13}) with $\delta\mathcal{R(\omega)}$ in
the stationary quasiparticle basis. To use it in the stationary code it has
to be transformed back to the original single-particle basis
\begin{equation}
\delta\mathcal{R(\omega)}=\left(
\begin{array}
[c]{cc}%
\delta\rho & \delta\kappa\\
-\delta\bar{\kappa}^{\ast} & -\delta\rho^{\ast}%
\end{array}
\right)  =\mathcal{W}\left(
\begin{array}
[c]{cc}%
0 & X\\
Y & 0
\end{array}
\right)  \mathcal{W}^{\dag} \;, 
\end{equation}
and one finds 
\begin{eqnarray}
\delta\rho &=& UXV^{\intercal}+V^{\ast}YU^{\dag},\\
\delta\kappa &=& UXU^{\intercal}+V^{\ast}YV^{\dag},\\
\delta\bar{\kappa}^{\ast} &=& -U^{\ast}YU^{\dag}-VXV^{\intercal}\;.
\end{eqnarray}
In this basis we derive the matrix elements of $\delta
\mathcal{H(\omega)}$ in Eq. (\ref{E18})%
\begin{eqnarray}
\delta h &=& \lim_{\eta\rightarrow0}\frac{1}{\eta}(h(\rho_{0}+\delta
\rho)-h(\rho_{0})), \label{eq:der-ham}\\
\delta\Delta &=& \lim_{\eta\rightarrow0}\frac{1}{\eta}(\Delta(\kappa
_{0}+\delta\kappa)-\Delta(\kappa_{0})),\label{eq:der-delta}\\
\delta\bar{\Delta} &=& \lim_{\eta\rightarrow0}\frac{1}{\eta}(\Delta(\kappa
_{0}+\delta\bar{\kappa})-\Delta(\kappa_{0}))\label{eq:der-delta_bar}\;,
\end{eqnarray}
and $\delta H^{20}(\omega)$ and $\delta H^{02}(\omega)$ are obtained by
transforming back to the quasiparticle basis
\begin{equation}
\delta\mathcal{\bar{H}(\omega)}=\left(
\begin{array}
[c]{cc}%
U^{\dag} & V^{\dag}\\
V^{\intercal} & U^{\intercal}%
\end{array}
\right)  \left(
\begin{array}
[c]{cc}%
\delta h & \delta\Delta\\
-\delta\bar{\Delta}^{\ast} & -\delta h^{\intercal}%
\end{array}
\right)  \left(
\begin{array}
[c]{cc}%
U & V^{\ast}\\
V & U^{\ast}%
\end{array}
\right)\;.
\end{equation}
The explicit expressions for $\delta H^{20}$ and $\delta
H^{02}$ read
\begin{eqnarray}
\delta H^{20}(\omega) &=& U^{\dag}\delta hV^{\ast}-V^{\dag}\delta
h^{\intercal}U^{\ast}+U^{\dag}\delta\Delta U^{\ast}-V^{\dag}\delta\bar{\Delta
}^{\ast}V^{\ast}\\
\delta H^{02}(\omega) &=& V^{\intercal}\delta hU-U^{\intercal}\delta
h^{\intercal}U+V^{\intercal}\delta\Delta V-U^{\intercal}\delta\bar{\Delta
}^{\ast}U \;.
\end{eqnarray}
Eqs. (\ref{E16}) and (\ref{E17}) are solved iteratively using the
Broyden method \cite{Sto.11}, and the transition density for each particular frequency 
$\omega$ reads%
\begin{equation}
\delta\rho_{tr}(\mathbf{r)}=-\frac{1}{\pi}\operatorname{Im}\delta
\rho(\mathbf{r}).
\end{equation}
The transition strength is calculated from%
\begin{equation}
S(f,\omega\mathbf{)}=-\frac{1}{\pi}\operatorname{Im}\text{Tr}[f(UXV^{\intercal
}+V^{\ast}YU^{\dag})]\; ,
\end{equation}
and in the present study we only consider isoscalar monopole transitions induced
by the single-particle operator%
\begin{equation}
f=\sum_{i=1}^{A}r_{i}^{2}.
\end{equation}
 

\section{\label{sec-numerical} Numerical implementation and test calculations}

The FAM for the relativistic QRPA is implemented using the stationary RHB code 
in which the single-nucleon Hartree-Bogoliubov equation (\ref{eq:RHB}) is 
solved by expanding the Dirac spinors in terms of eigenfunctions of an 
axially symmetric harmonic oscillator potential (cf. Appendix \ref{Sec-App1}). 
The expressions for the matrix elements of the monopole operator in this basis 
are given in Appendix \ref{Sec-App2}.

In the present illustrative study we employ the relativistic functional DD-PC1 \cite{NVR.08}. 
Starting from microscopic nucleon self-energies in nuclear matter, and empirical 
global properties of the nuclear matter
equation of state, the coupling parameters of DD-PC1 were fine-tuned to the experimental
masses of a set of 64 deformed nuclei in the mass regions $A \approx 150-180$ and
$A \approx 230-250$. The functional has been further tested in calculations of ground-state 
properties of medium-heavy and heavy nuclei, including binding energies, charge radii, 
deformation parameters, neutron skin thickness, and excitation energies of giant multipole 
resonances. A pairing force separable in momentum space \cite{TMR.09a}: 
$\bra{k}V^{^1S_0}\ket{k'} = - G p(k) p(k')$ will be here used in the \textit{pp} channel. 
By assuming a simple Gaussian ansatz $p(k) = e^{-a^2k^2}$, the two parameters $G$ and $a$ 
were adjusted to reproduce the density dependence of the gap at the Fermi surface 
in nuclear matter, calculated with the pairing part of the Gogny interaction. When
transformed from momentum to coordinate space, the interaction takes the form: 
\begin{equation}
V(\bff{r}_1,\bff{r}_2,\bff{r}_1^\prime,\bff{r}_2^\prime)
=G\delta \left(\bff{R}-\bff{R}^\prime \right)P(\bff{r})P(\bff{r}^\prime)
\frac{1}{2}\left(1-P^\sigma \right),
\label{pp-force}
\end{equation}
where $\bff{R}=\frac{1}{2}\left(\bff{r}_1+\bff{r}_2\right)$ and 
$\bff{r}=\bff{r}_1-\bff{r}_2$ denote the center-of-mass and the relative coordinates, 
respectively, and $P(\bff{r})$ is the Fourier transform of $p(k)$:
$P(\bff{r}) = 1/\left(4\pi a^2 \right)^{3/2}e^{-\bff{r}^2/4a^2}$.
The actual implementation of the FAM does not, of course, depend on the 
choice of the relativistic density functional or the pairing functional. 

To avoid the occurrence of singularities in the right-hand side of Eqs.~(\ref{E16}) and (\ref{E17}), 
the frequency $\omega$ is replaced by $\omega + i \gamma$
with a small parameter $\gamma$, related to the Lorentzian
smearing $\Gamma = 2\gamma$  in RQRPA calculations. 
Eqs.~(\ref{E16}) and (\ref{E17}) are solved iteratively. The solution is reached when the maximal
difference between collective amplitudes corresponding to two successive iterations decreases below a
chosen threshold ($\epsilon =10^{-6}$).  The stability and rapid convergence
of the FAM iteration procedure is ensured by adopting the modified Broyden's 
procedure~\cite{Johnson.88,Baran.08}, which is also implemented in the calculation
of the RHB equilibrium solution. Compared to ground state calculations, the
use of Broyden's method in the FAM for QRPA requires an increase of the number of vectors retained
in Broyden's history ($M=20$ for the FAM, compared to $M=7$ for the RHB). With this modification 
FAM solutions have been achieved with less than 40 iterations for all examples considered in the 
present illustrative calculations. The FAM for QRPA necessitates the inclusion of time-odd 
terms (currents) in the calculation of induced fields (cf. Appendix \ref{Sec-App3}). 
The FAM equations for the case of time-reversal, reflection and axial symmetries are 
detailed in Appendix \ref{Sec-App4}.

\begin{figure}[htb!]
\centering
\includegraphics[width=\textwidth]{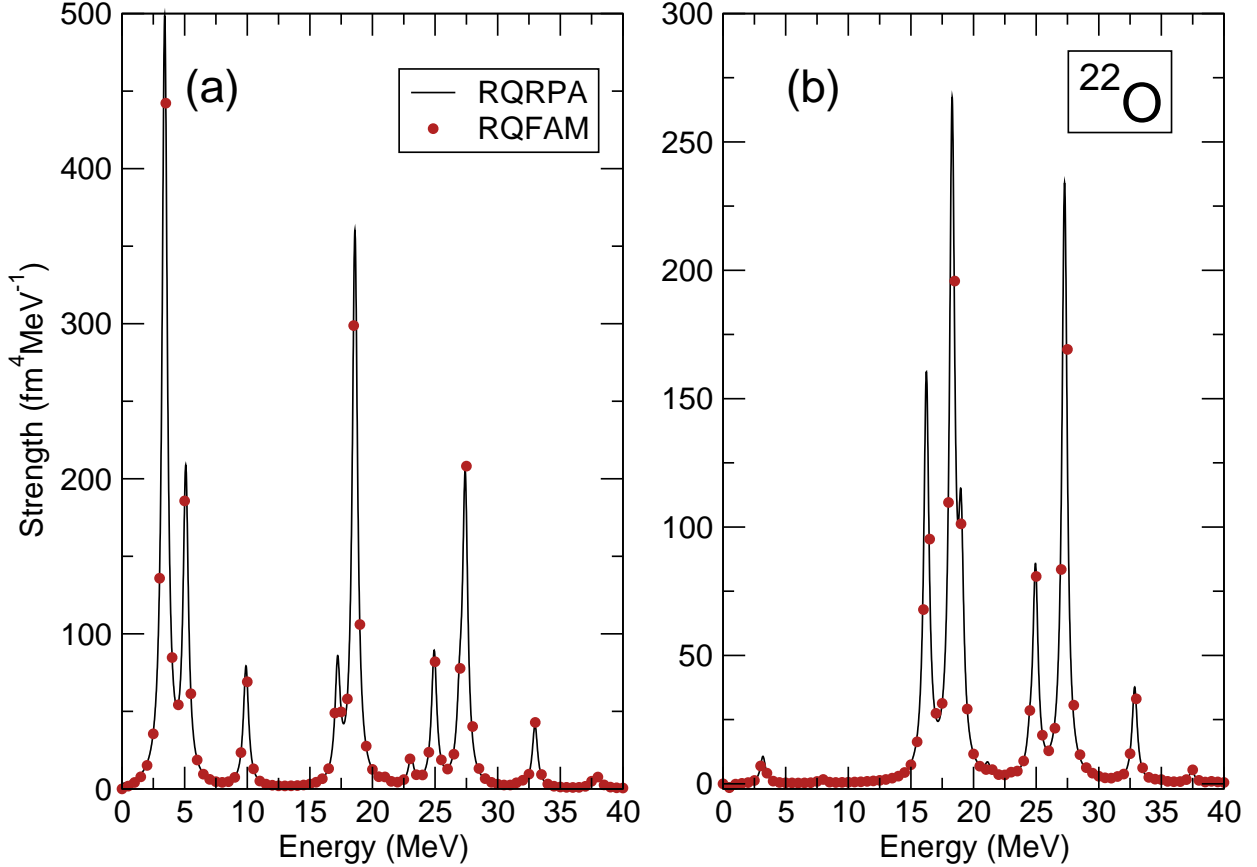}
\caption{\label{fig:numerical1}(Color online) Strength functions of the isoscalar monopole
operator for $^{22}$O. The solid curves denote the RQRPA response, FAM results are
indicated by (red) symbols. The two panels correspond to a calculation without 
dynamical pairing (panel (a)), and to a fully self-consistent calculation with pairing included 
in the RQRPA residual interaction and FAM induced fields (panel (b)). The single-nucleon 
wave functions are expanded in a basis of 10 oscillator shells, and the response is 
smeared with a Lorentzian of $\Gamma = 2\gamma = 0.5$ MeV width.}
\end{figure}

To verify the numerical implementation and accuracy of our FAM model, 
a simple test calculation has been performed for the light spherical nucleus $^{22}$O. 
In this case we could directly compare the FAM results to those obtained 
using the standard computer code for the RQRPA matrix~\cite{Paar2003}. This comparison 
presents an excellent test of both codes because the FAM formalism employs 
only numerical derivatives of the single-particle Hamiltonian and the pairing field, 
whereas the QRPA codes uses explicit expressions for the matrix 
elements of the residual interaction.
In Fig.~\ref{fig:numerical1} we display the isoscalar strength functions of the 
monopole operator $\sum_{i=1}^A{r_i^2}$ for $^{22}$O. The 
panel (a) corresponds to a calculation without dynamical pairing, that is, pairing is only 
included in the calculation of the RHB ground state but not in the residual 
interaction (QRPA) or induced fields (FAM). The strength functions in the 
panel (b) are calculated fully self-consistently with dynamical pairing. In both panels 
the solid curves denote the RQRPA response, whereas symbols correspond to the 
FAM results. Firstly we note that in both cases the RQRPA and FAM results 
coincide exactly at all excitation energies. In the calculation without dynamical 
pairing, that is, by including pairing correlations only in the RHB
ground state, one notices the occurrence of a strong spurious response below 10 MeV.  
This Nambu-Goldstone mode is driven to approximately zero excitation 
energy (in this particular calculation 
it is located below 0.2 MeV) when pairing correlations are consistently included 
in the QRPA residual interaction and FAM induced fields.

\begin{figure}[htb!]
\centering
\includegraphics[width=\textwidth]{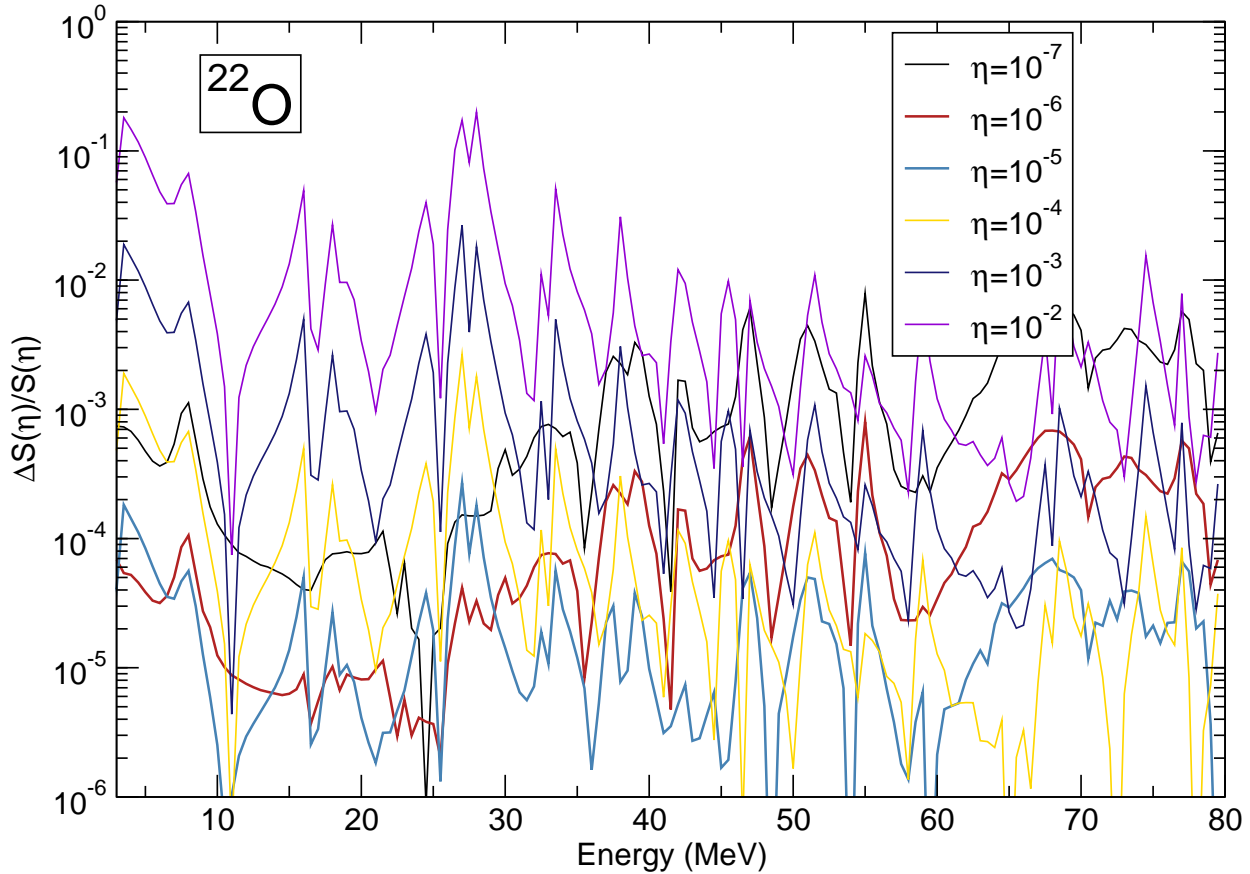}
\caption{\label{fig:numerical2}(Color online) The relative accuracy of the strength function for 
the isoscalar monopole operator in $^{22}$O (see Eq.~(\ref{eq:relative-accuracy})). 
The curves are plotted for several values of the
parameter $\eta$  and span a broad interval of excitation energies.}
\end{figure}

Fig.~\ref{fig:numerical2} shows the stability of the current implementation of the 
FAM method for a broad range of values of 
the parameter $\eta$ that is used to calculate the numerical derivatives in Eqs.~(\ref{eq:der-ham})
-- (\ref{eq:der-delta_bar}).
The relative accuracy of the strength function is defined as
\begin{equation}
\label{eq:relative-accuracy}
\frac{\Delta S(\omega,\eta)}{S(\omega,\eta)} = \frac{1}{S(\omega,\eta)}|S(\omega,10\eta)-S(\omega,\eta)|.
\end{equation}
In practice the accuracy can only be improved by reducing $\eta$ down to $10^{-6}$. A further 
decrease of this parameter introduces numerical noise which deteriorates the accuracy 
of the FAM method, and thus $\eta=10^{-6}$ has been used throughout this study.

\section{\label{sec-illustrative-calculations} Illustrative calculations: samarium isotopes}

Collective nucleonic oscillations along different axes in deformed nuclei and mixing 
of different modes lead to a broadening and splitting of giant resonance structures \cite{HW.01}. 
The giant dipole resonance (GDR), for instance, displays a 
two-component structure in deformed nuclei and the origin of this splitting 
are the different frequencies of oscillations along the major and minor axes.
In axially deformed nuclei 
the isoscalar giant quadrupole (ISGQR) resonance displays three components with
$K^\pi=0^+,1^+,2^+$ \cite{Kis.75,Miura.77}, where $K$ denotes the projection of   
the total angular momentum $I=2^+$ on the intrinsic symmetry axis. 
The isoscalar giant monopole resonance (ISGMR) in deformed nuclei
mixes with the $K^\pi=0^+$ component of the ISGQR  
and a two-peak structure of the monopole resonance is observed  \cite{Garg.80,Morsch.82}. 
In a recent study of the roles of deformation and neutron excess on the giant monopole
resonance in neutron-rich deformed Zr isotopes~\cite{Yos.10}, based on the deformed 
Skyrme -- HFB + QRPA model, the evolution of the 
two-peak structure of the ISGMR has been investigated. The theoretical analysis has shown 
that the lower peak is associated with the mixing between the ISGMR and the
$K^\pi=0^+$ component of the ISGQR, and the transition strength of the lower peak 
increases with neutron excess. Here we apply the FAM method for the relativistic QRPA to a  
calculation of the isoscalar $K^\pi=0^+$ 
strength functions in the chain of even-even Sm isotopes, starting from the neutron-deficient $^{132}$Sm 
isotope and extending to the neutron-rich $^{160}$Sm isotope. 
The calculations have been performed in the harmonic oscillator basis with $N_{max}=18$ 
oscillator shells for the upper component and $N_{max}=19$ shells for the lower 
component of the Dirac 
spinors~\cite{GTR.90}. It has been demonstrated in Ref.~\cite{Sto.11} that by using 
$N_{max}=18$ oscillator shell basis one obtains convergent results even for the superdeformed states.

\begin{figure}
\centering
\includegraphics[width=\textwidth]{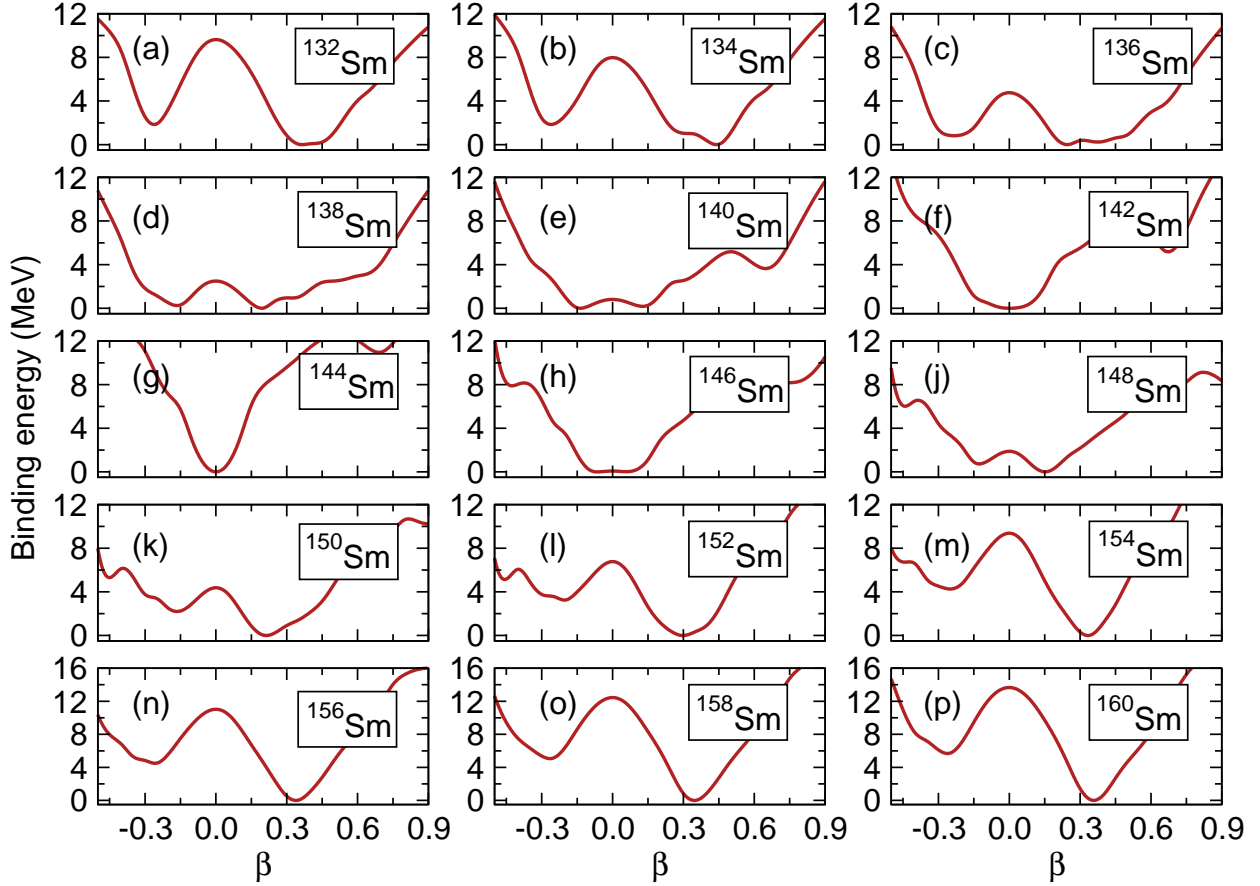}
\caption{\label{fig:pes_sm}(Color online) Self-consistent RHB binding energy curves of the 
even-even $^{132-160}$Sm isotopes as functions of the axial deformation parameter $\beta$.
Energies are normalized with respect to the binding energy of the absolute minimum for
each isotope.}
\end{figure}

Fig.~\ref{fig:pes_sm} displays the energy curves of Sm isotopes calculated with 
the constraint on the axial quadrupole moment, as functions of the axial deformation parameter $\beta$. 
Energies are normalized with respect to the binding energy of the 
absolute minimum for each isotope. For the isotopes with a prolate equilibrium deformation
($^{132-136}$Sm and $^{152-160}$Sm), an additional minimum is predicted on the oblate side
and  the two minima are separated by a potential barrier. 
In neutron-deficient isotopes both the oblate minimum and the
potential barrier are considerably lower compared to the neutron-rich nuclei. Both
$^{136}$Sm and $^{152}$Sm, that is, nuclides at the borders of the region of weakly 
deformed and/or spherical systems around the neutron shell-closure at $N=82$, 
exhibit soft potentials with wide minima on the prolate side.
$^{138-150}$Sm display two weakly deformed
and almost degenerated minima, and the isotopes $^{142,144}$Sm are spherical.

\begin{figure}
\centering
\includegraphics[width=\textwidth]{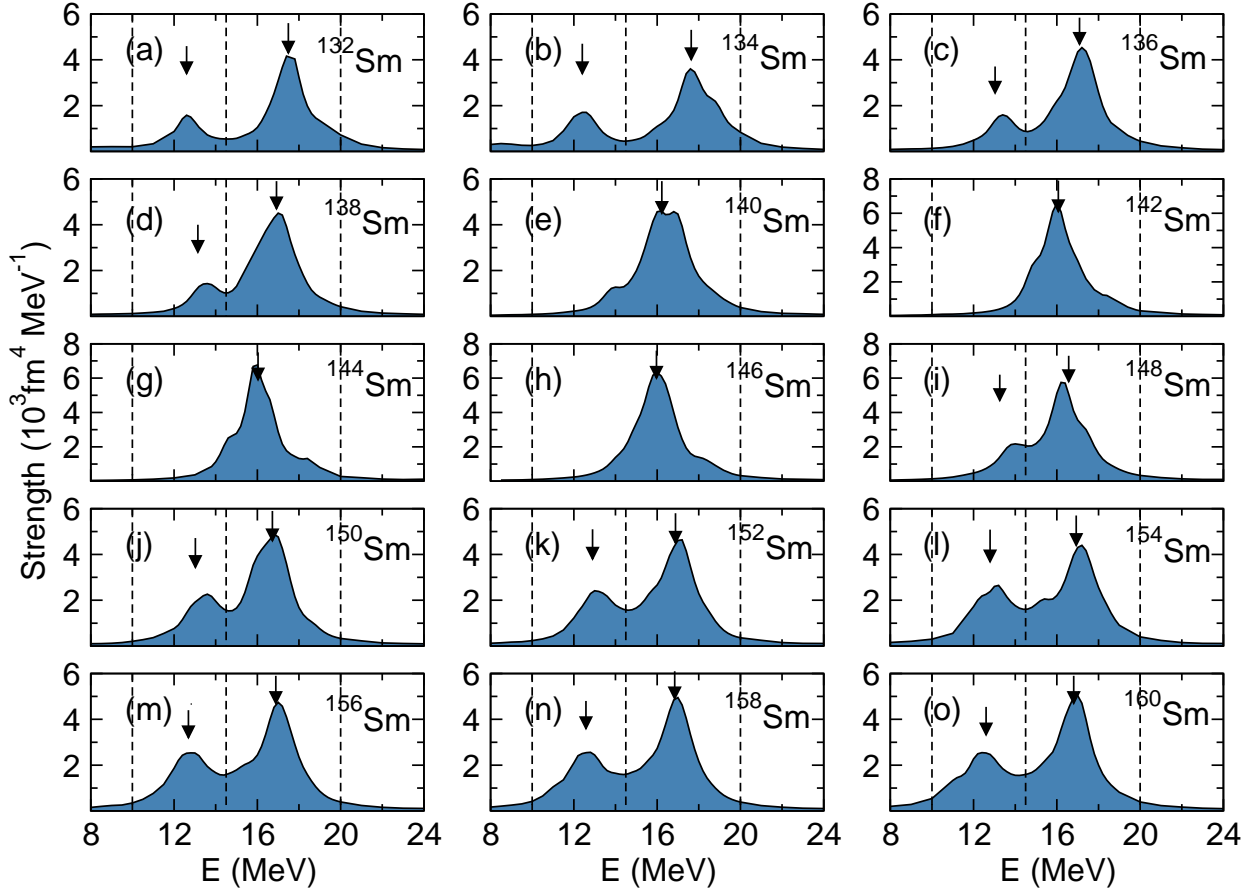}
\caption{\label{fig:sm-response}(Color online) Evolution of the $K^\pi=0^+$ strength functions 
in $^{132-160}$Sm. The arrows indicate the positions of the mean energies $m_1/m_0$
calculated in the energy intervals $10< E < 14.5$ MeV and $14.5 < E < 20$ MeV.}
\end{figure}

For each isotope in the chain $^{132-160}$Sm the calculated $K^\pi=0^+$ response is 
shown in Fig.~\ref{fig:sm-response}. The principal result is the 
splitting of the $K^\pi=0^+$ strength into two 
peaks for the deformed isotopes. The arrows indicate the positions of the mean energies $m_1/m_0$,
that is, the ratio of the energy-weighted sum (EWS) and the non-energy-weighted sum, calculated 
in the energy intervals $10 < E < 14.5$ MeV for the low-energy (LE) peak, and $14.5  < E < 20$ MeV for 
the high-energy (HE) peak. 
The HE peak of the monopole strength distribution is located slightly above the
energy of the ISGMR in the spherical isotope $^{144}$Sm, whereas the LE peak
appears in the energy region where the giant quadrupole
resonance in $^{144}$Sm is located ($E_{\textnormal{ISGQR}}=14$ MeV). 
With increasing deformation (cf. Fig.~\ref{fig:pes_sm}) the HE peak is shifted to 
higher energy because of the coupling with the $K^\pi=0^+$ component of the ISGQR, 
and the LE peak is simultaneously lowered in energy. It should be noted that the $K^\pi=0^+$ components
of other resonances also contribute to the LE and HE peaks, but to 
a much lesser extent.

\begin{figure}
\centering
\includegraphics[width=0.8\textwidth]{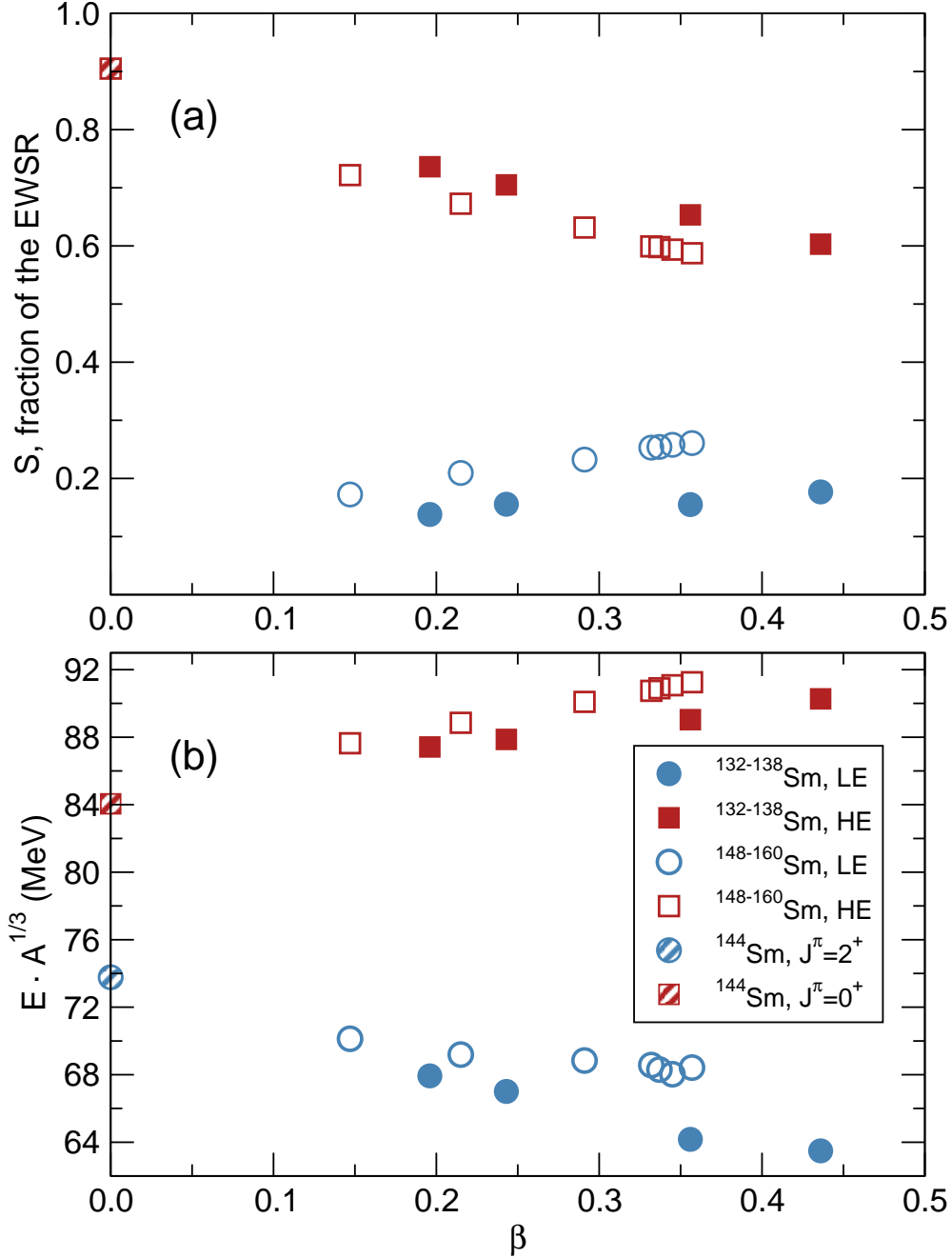}
\caption{\label{fig:sm-centroids}(Color online) Panel (a): 
fraction of the EWSR for the HE and LE  
components of the $K^\pi=0^+$ strength in the deformed nuclei $^{132-138}$Sm and $^{148-160}$Sm, 
calculated in the energy intervals 
$14.5 < E < 20$ MeV and $10<E<14.5$ MeV, respectively, plotted as functions of the 
equilibrium deformation parameter. 
Panel (b): the corresponding mean energies $m_1/m_0$ of the HE and LE peaks, 
denoted by squares and circles, respectively, as functions of the equilibrium value of $\beta$.}
\end{figure}

In the panel (b) of Fig.~\ref{fig:sm-centroids} we display the mean energies of the 
HE (squares) and LE (circles)
peaks as functions of the equilibrium deformation parameter $\beta$.
The calculated energies are multiplied by the factor $A^{1/3}$ to account for the 
empirical mass dependence of the ISGMR excitation energy $E \sim A^{-1/3}$. 
With increasing equilibrium deformation the splitting between the LE and HE components 
becomes larger, although the trend is not quite the same for the isotopes with $A<144$
and $A>144$. This difference can be caused by shell effects or different 
neutron to proton ratio.
The fractions of the energy weighted sum-rule (EWSR) for the HE and LE
energy peaks are shown in the panel (a) of Fig.~\ref{fig:sm-centroids}.
The fractions of the EWSR are calculated in the same intervals as the mean energies: 
$10 < E < 14.5$ MeV for the low-energy (LE) component, and $14.5  < E < 20$ MeV for 
the high-energy (HE) region. Generally the fraction of the EWSR in the LE mode increases
with deformation, but again the trend is slightly different for for the isotopes with $A<144$
and $A>144$. 
The sum of the LE and HE component amounts to about 90\% of the EWSR because the integration interval is limited to 20MeV. We have verified that by extending the integration limit to 50MeV over 99\% of the EWSR is exhausted. 
It should be noted that in a relativistic (Q)RPA the nonrelativistic sum rules are only approximately exhausted when the integration is performed only over positive 
energies~\cite{PW.85,SRM.89,McNeil.89}.
\begin{figure}
\centering
\includegraphics[width=\textwidth]{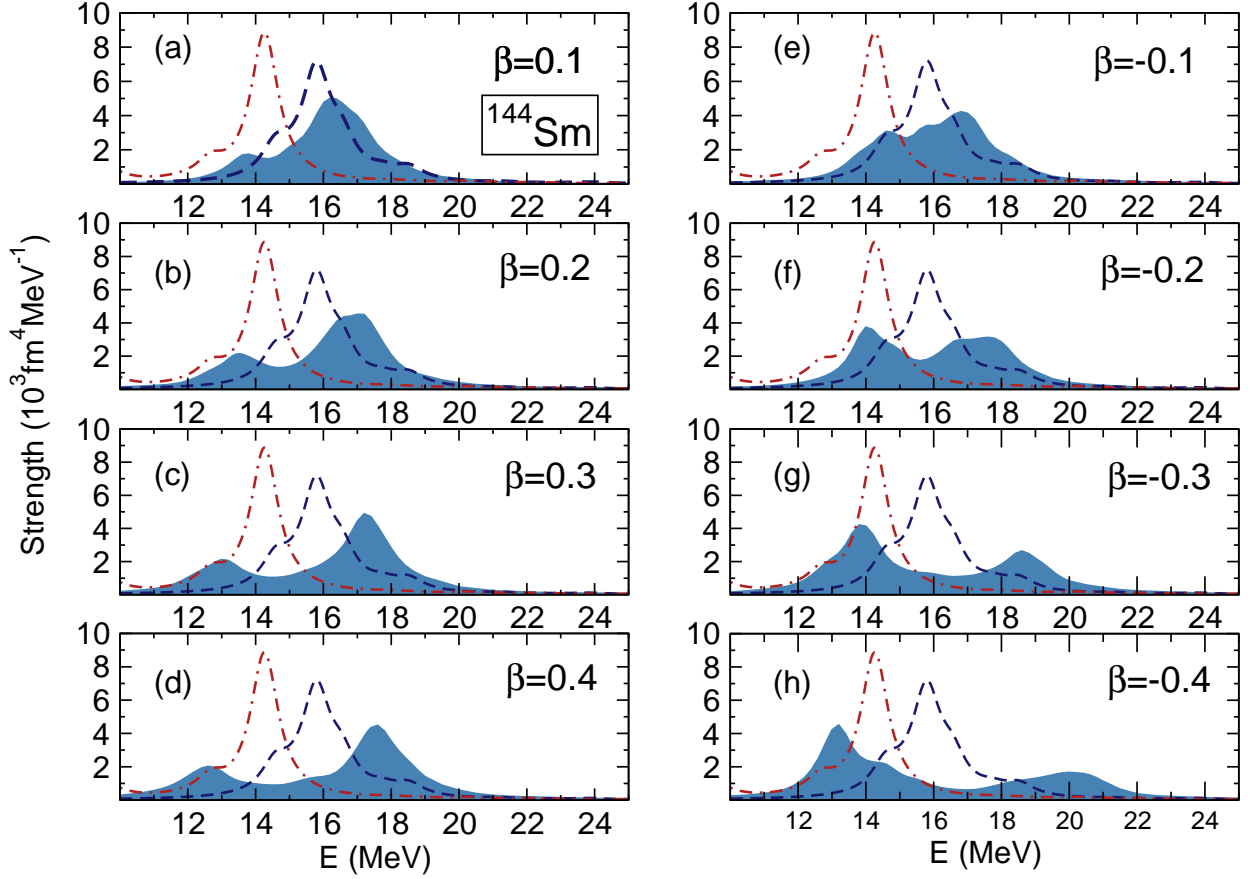}
\caption{\label{fig:sm144-cstr}(Color online) The $K^\pi=0^+$ strength distributions in 
$^{144}$Sm for eight different values of the axial quadrupole constraint. 
The blue dashed and red dot-dashed curves denote the
$J=0^+$ and $J=2^+$ strengths for the $^{144}$Sm equilibrium spherical configuration, respectively. }
\end{figure}
The splitting of the $K^\pi=0^+$ strength in deformed systems can be studied in more detail by performing 
a deformation-constrained calculation for a single isotope. In Fig.~\ref{fig:sm144-cstr} we show 
the $K^\pi=0^+$ strength distributions in $^{144}$Sm isotope for eight different values of the 
axial quadrupole constraint, from $\beta=-0.4$ to $\beta=0.4$. 
The blue dashed and red dot-dashed curves correspond to the monopole and quadrupole strength distributions for 
the equilibrium spherical configuration of $^{144}$Sm, respectively. Both for the prolate and oblate 
constrained configurations the splitting between the LE
and HE components of the $K^\pi=0^+$ strength increases with deformation. An interesting 
result is that the HE component of the $K^\pi=0^+$ strength distribution is
more pronounced for prolate configurations, whereas for oblate configurations the LE component 
becomes dominant.

\begin{figure}
\centering
\includegraphics[width=\textwidth]{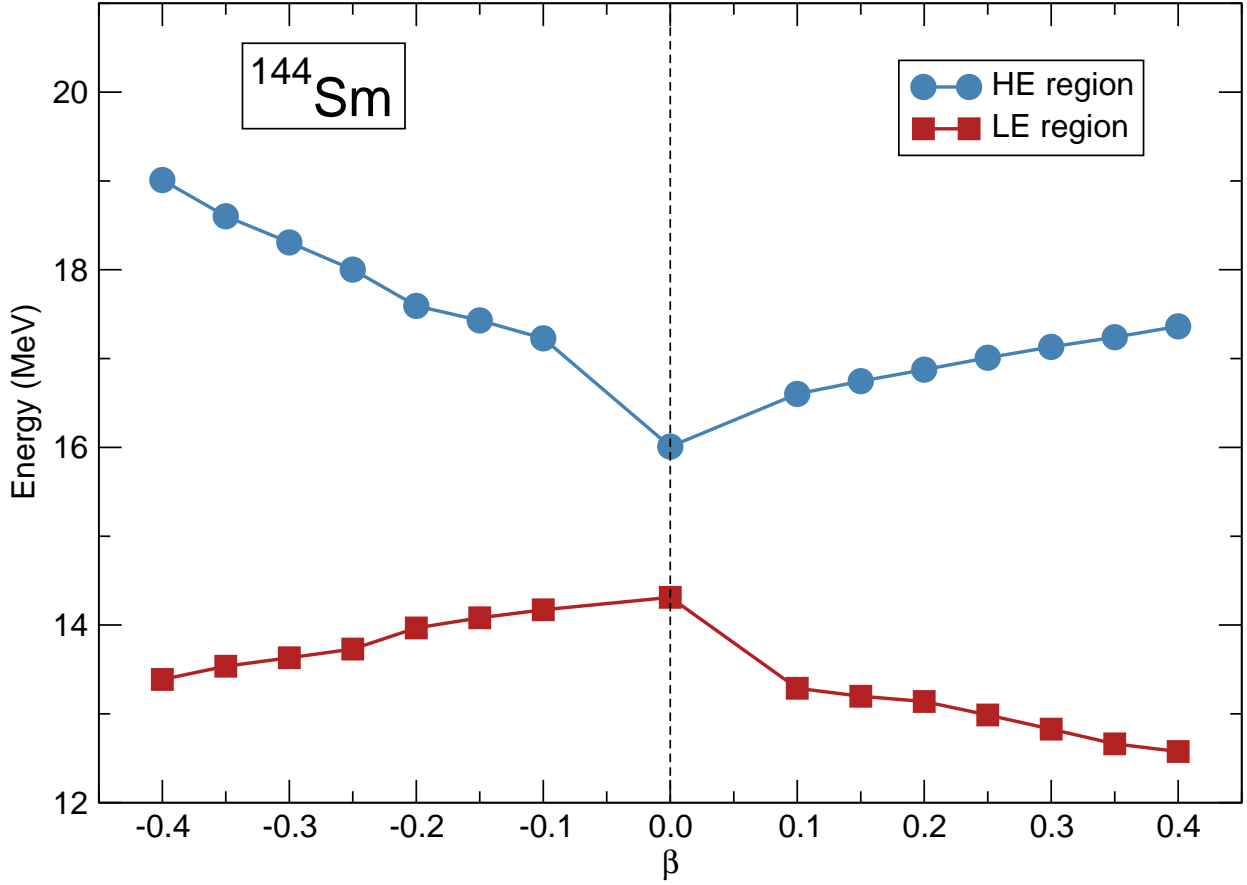}
\caption{\label{fig:evolution-sm144}(Color online) The mean energies  
$m_1/m_0$  of the HE and LE  
components of the $K^\pi=0^+$ strength distribution in $^{144}$Sm,
as functions of the constrained quadrupole deformation $\beta$.}
\end{figure}

Fig.~\ref{fig:evolution-sm144} compares the mean energies, that is, the  
ratio of the energy-weighted sum (EWS) and the non-energy-weighted sum $m_1/m_0$ 
of the HE and LE components of the $K^\pi=0^+$ strength distribution in $^{144}$Sm,
as functions of the constrained quadrupole deformation $\beta$. 
For the prolate configurations the moments of the strength distribution 
are calculated in the energy intervals  $10-14.5$ MeV (LE region)
and $14.5-20$ MeV (HE region). The corresponding intervals for oblate configurations are 
$10-15.5$ MeV (LE region) and $15.5-22.5$ MeV (HE region) (cf. Fig.~\ref{fig:sm144-cstr}). 
The repulsion between the LE and the HE components of the $K^\pi=0^+$ distribution is 
consistent with the result shown in Fig. 1 of Ref.~\cite{NA.85}. We note that the monopole ($I = 0^+$) 
and quadrupole ($I = 2^+$) strength distributions,
calculated for the spherical equilibrium configuration of $^{144}$Sm, are somewhat fragmented 
and this leads to the broadening of the $K^\pi=0^+$ strength distribution for deformed configurations
since each monopole state couples to each quadrupole state.

\begin{figure}
\centering
\begin{tabular}{cc}
\includegraphics[trim=7cm 0cm 1.5cm 0cm, clip=true, width=5.5cm]{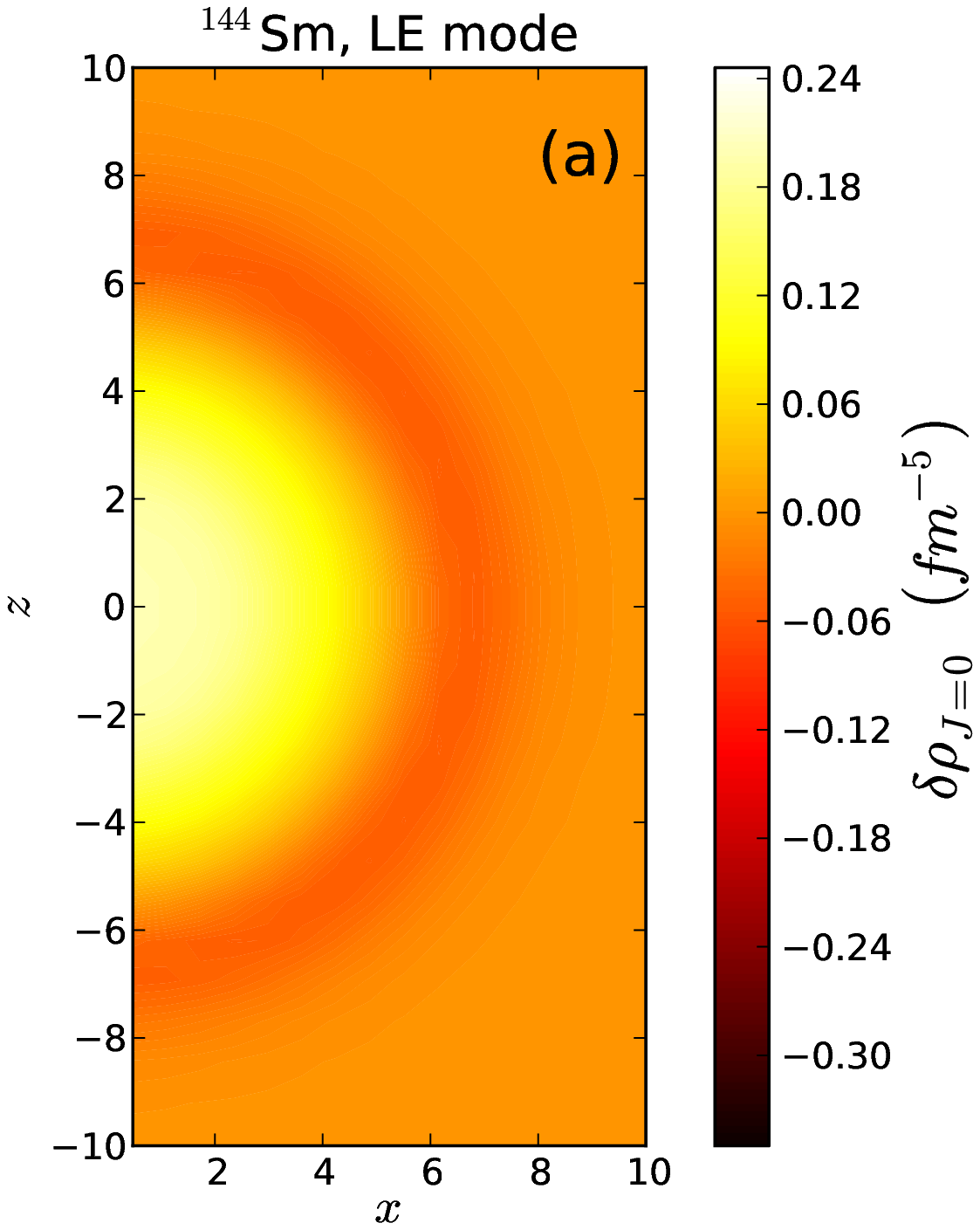} 
& \includegraphics[trim=7cm 0cm 1.5cm 0cm, clip=true,width=5.5cm]{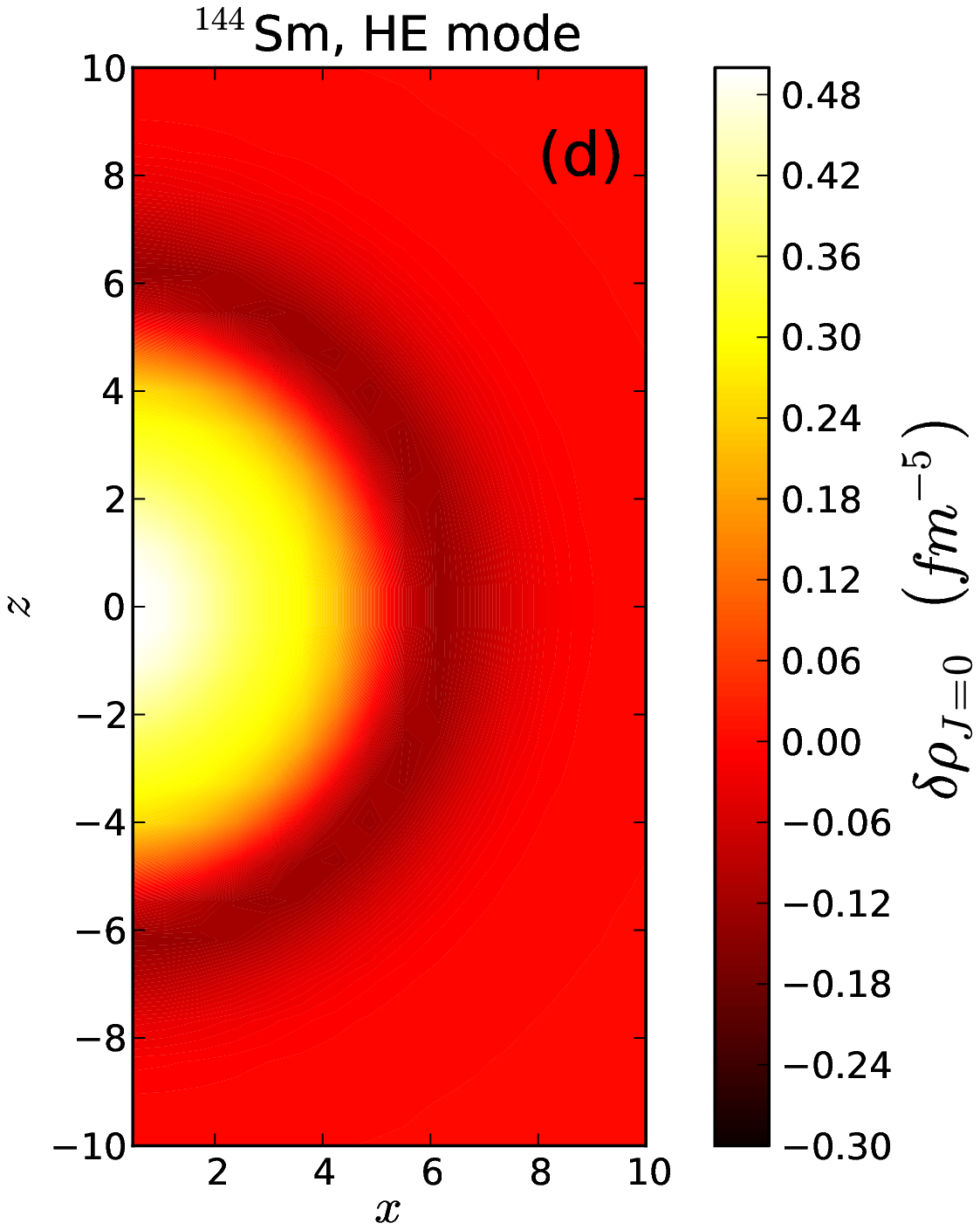} \\
\includegraphics[trim=7cm 0cm 1.5cm 1cm, clip=true, width=5.5cm]{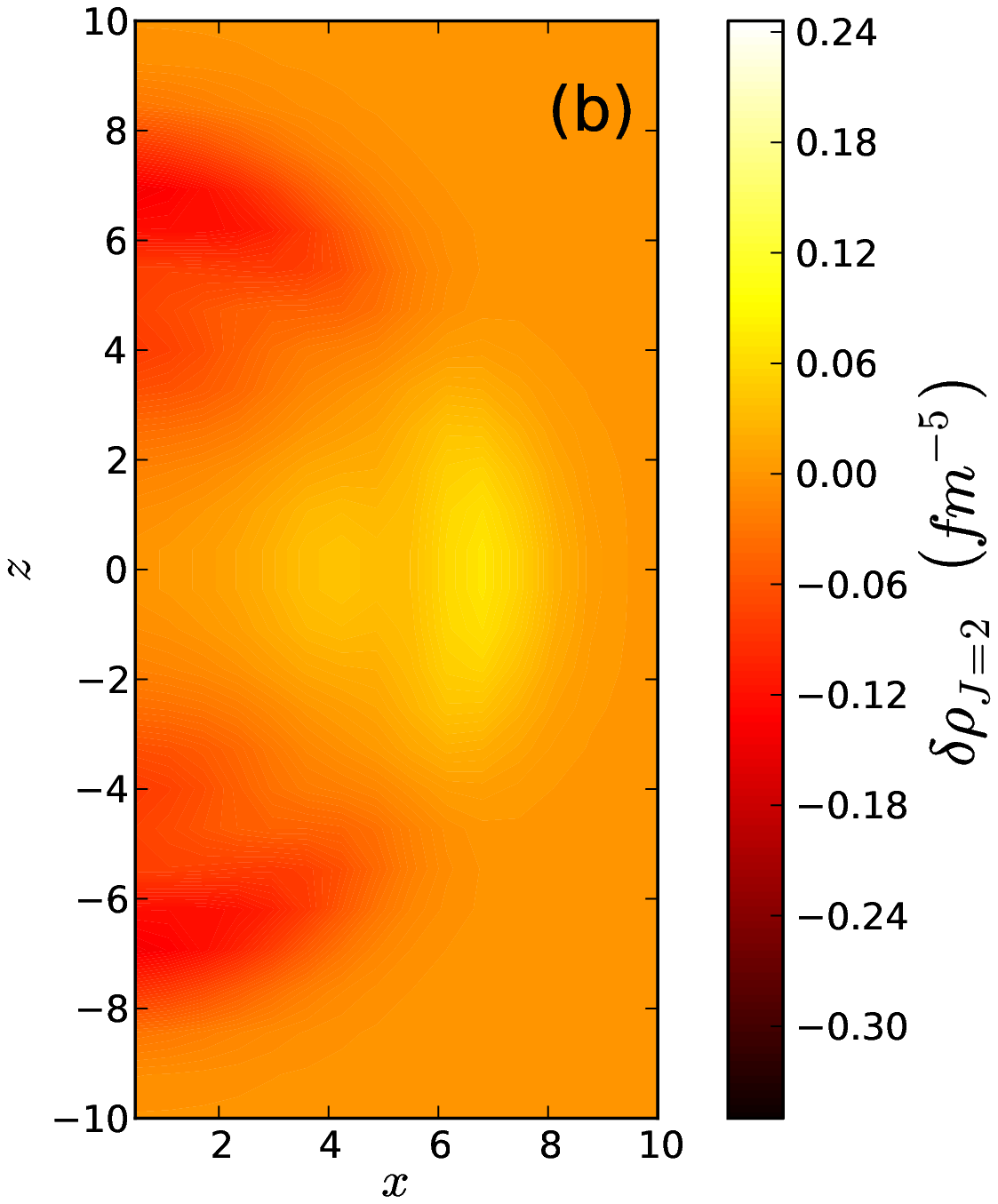} 
& \includegraphics[trim=7cm 0cm 1.5cm 1cm, clip=true, width=5.5cm]{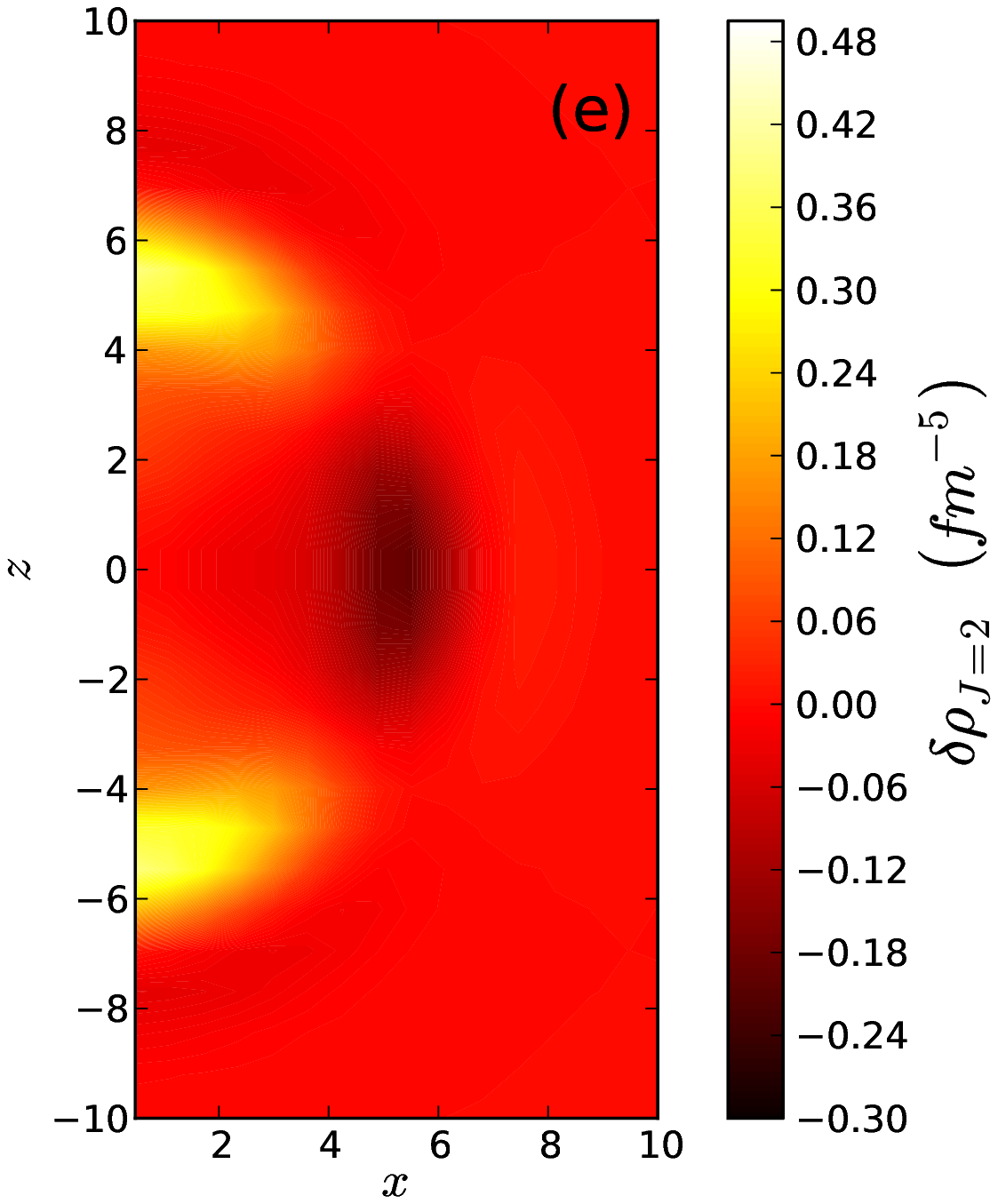} \\
\includegraphics[trim=7cm 0cm 1.5cm 1cm, clip=true, width=5.5cm]{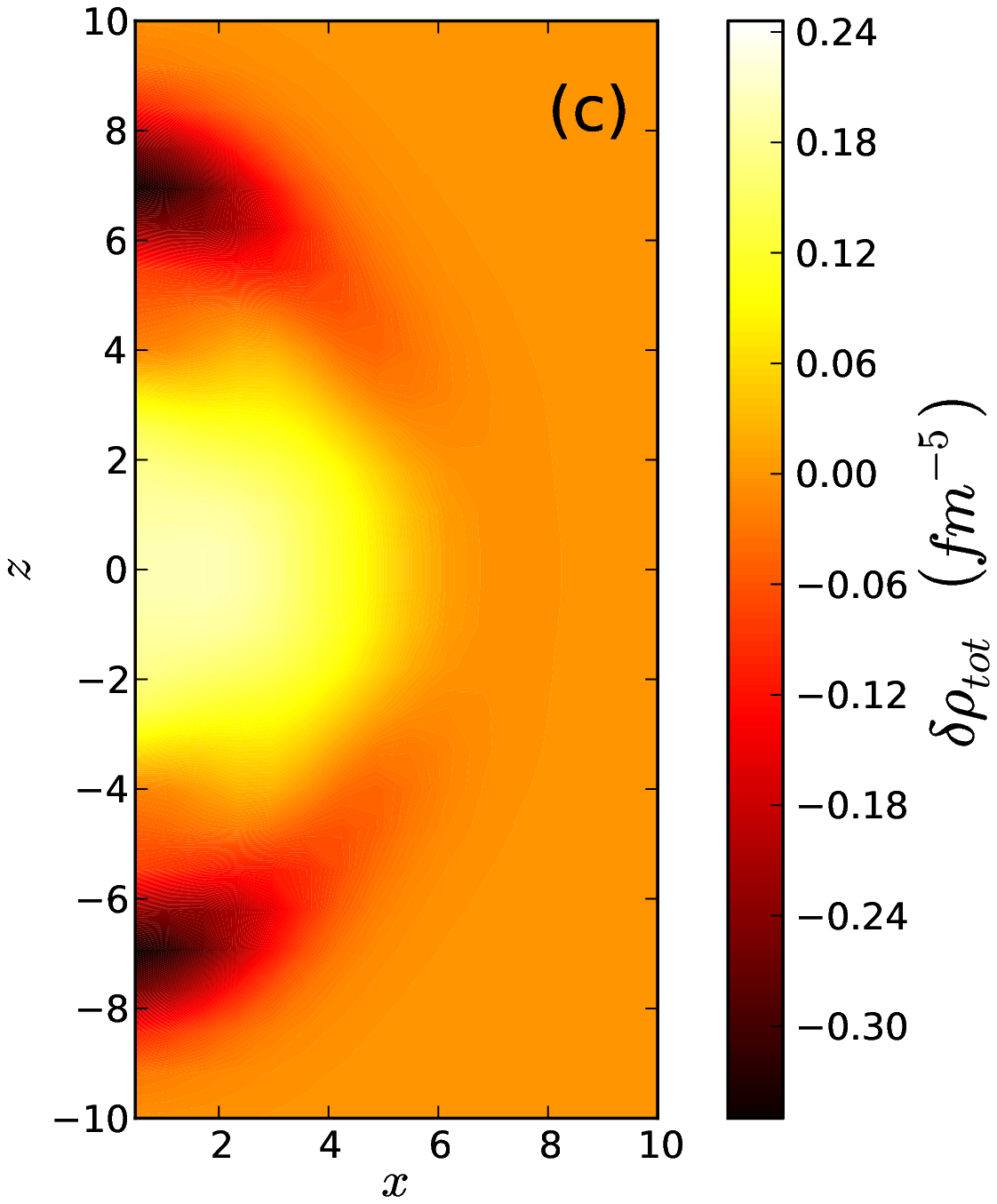} 
& \includegraphics[trim=7cm 0cm 1.5cm 1cm, clip=true, width=5.5cm]{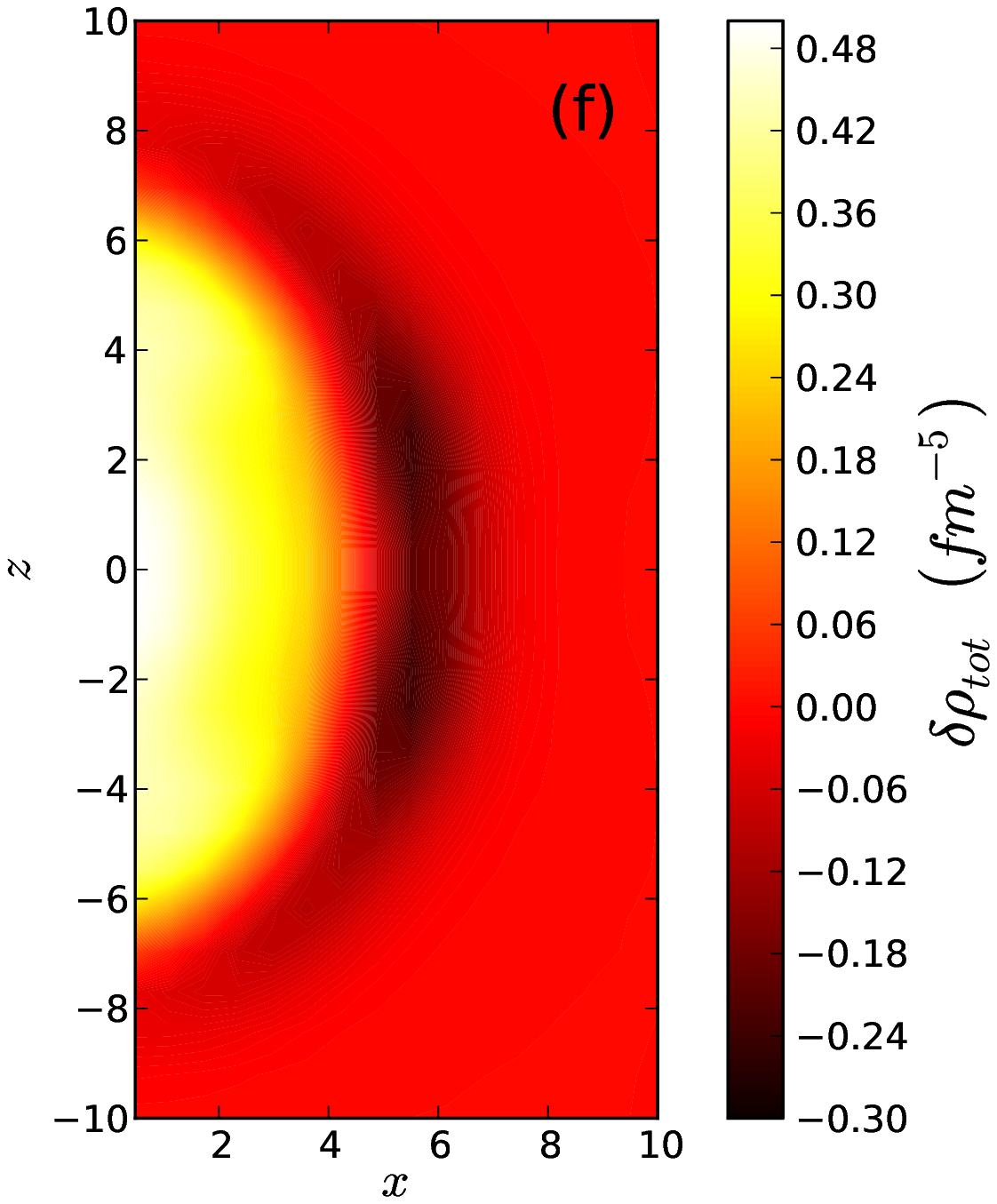}
\end{tabular}
\caption{\label{fig:sm-td-pro}(Color online) The $J=0$ and $J=2$ 
angular-momentum-projected transition densities, 
and the intrinsic transition densities for the LE (left column) and HE (right column) 
peaks of the ISGMR strength distribution in $^{144}$Sm. The stationary density corresponds 
to the prolate configuration with the constraint deformation $\beta=0.3$.
 }
\end{figure}

\begin{figure}
\centering
\begin{tabular}{cc}
\includegraphics[trim=7cm 0cm 1.5cm 0cm, clip=true, width=5.5cm]{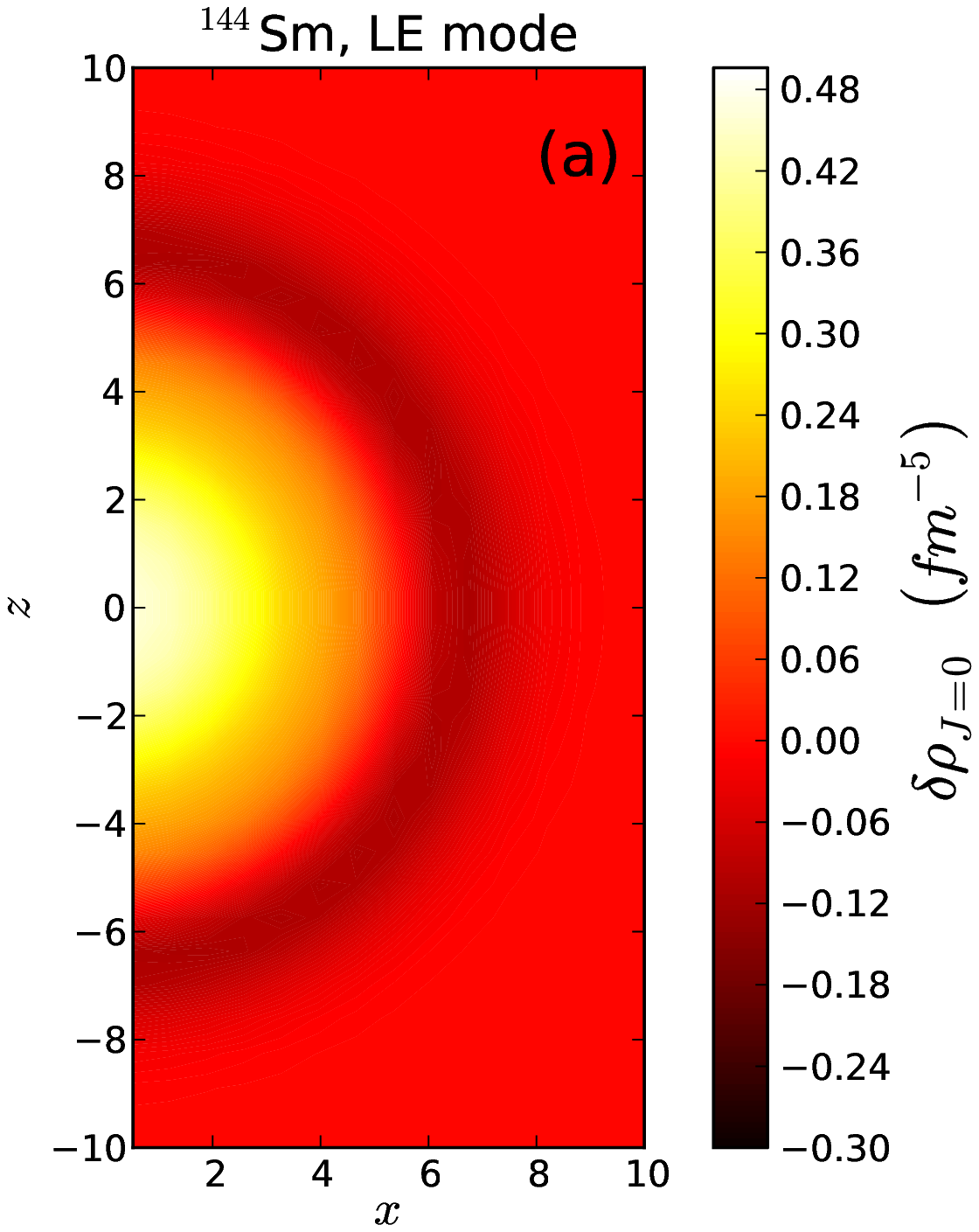} 
& \includegraphics[trim=7cm 0cm 1.5cm 0cm, clip=true, width=5.5cm]{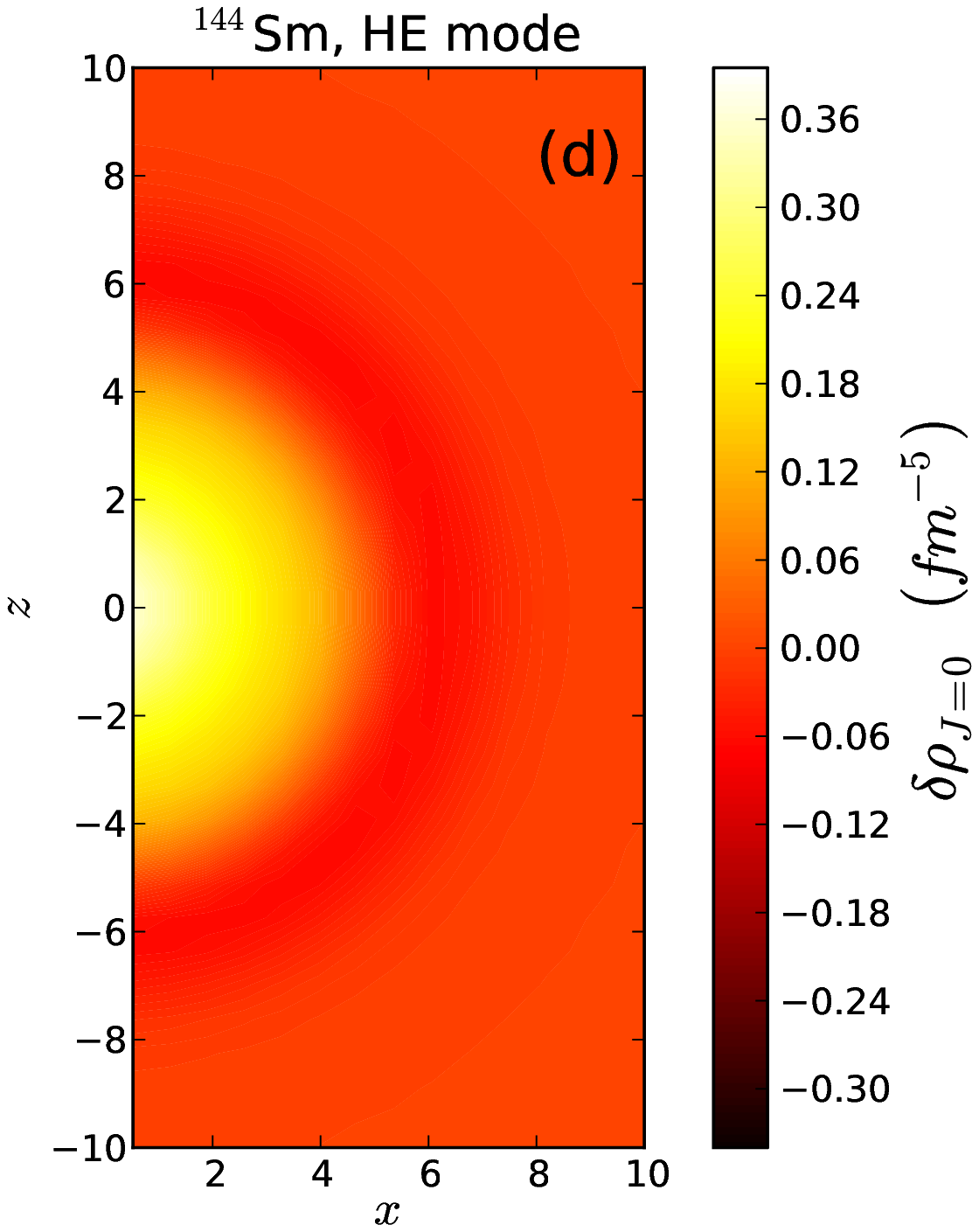} \\
\includegraphics[trim=7cm 0cm 1.5cm 1cm, clip=true, width=5.5cm]{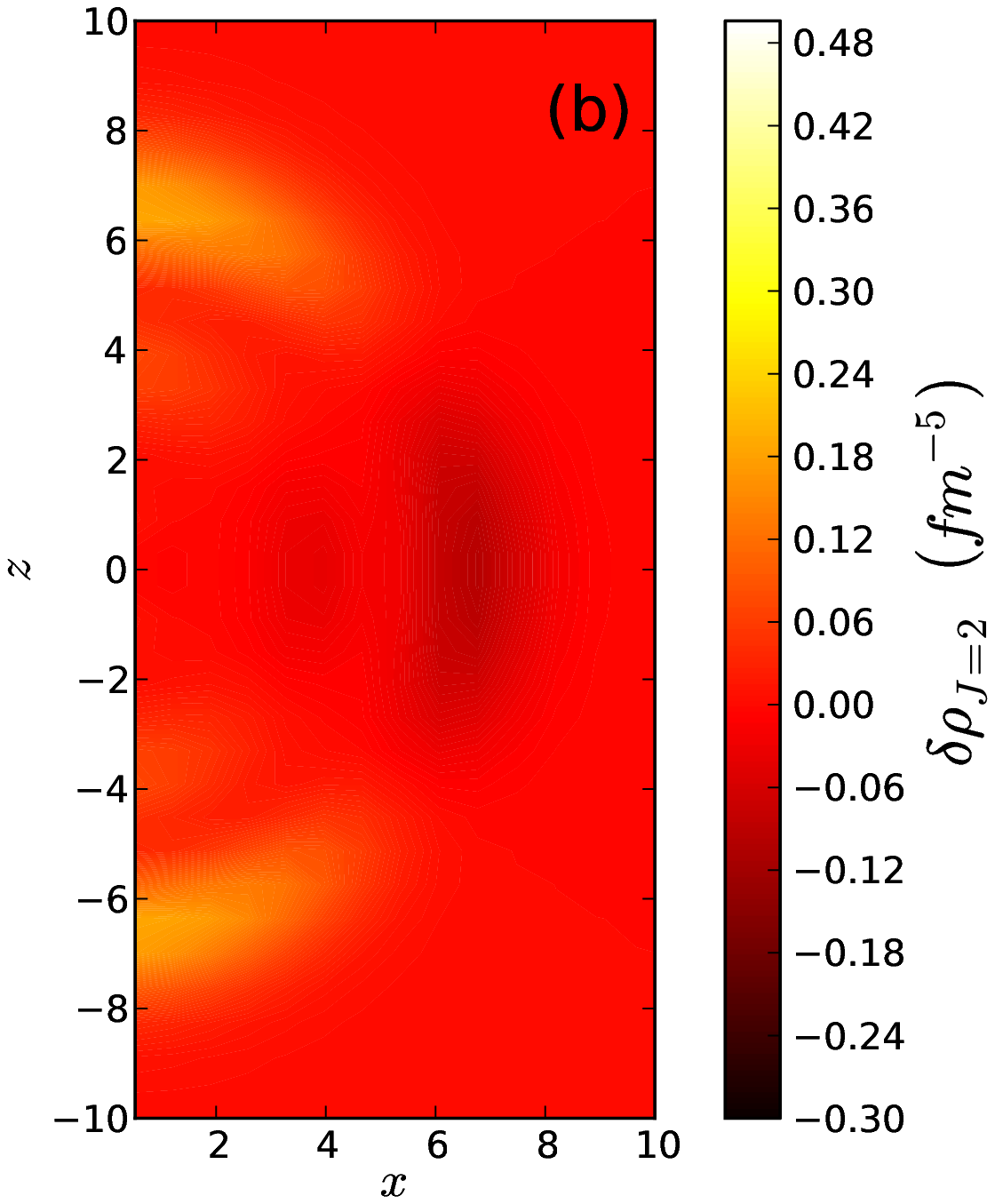} 
& \includegraphics[trim=7cm 0cm 1.5cm 1cm, clip=true, width=5.5cm]{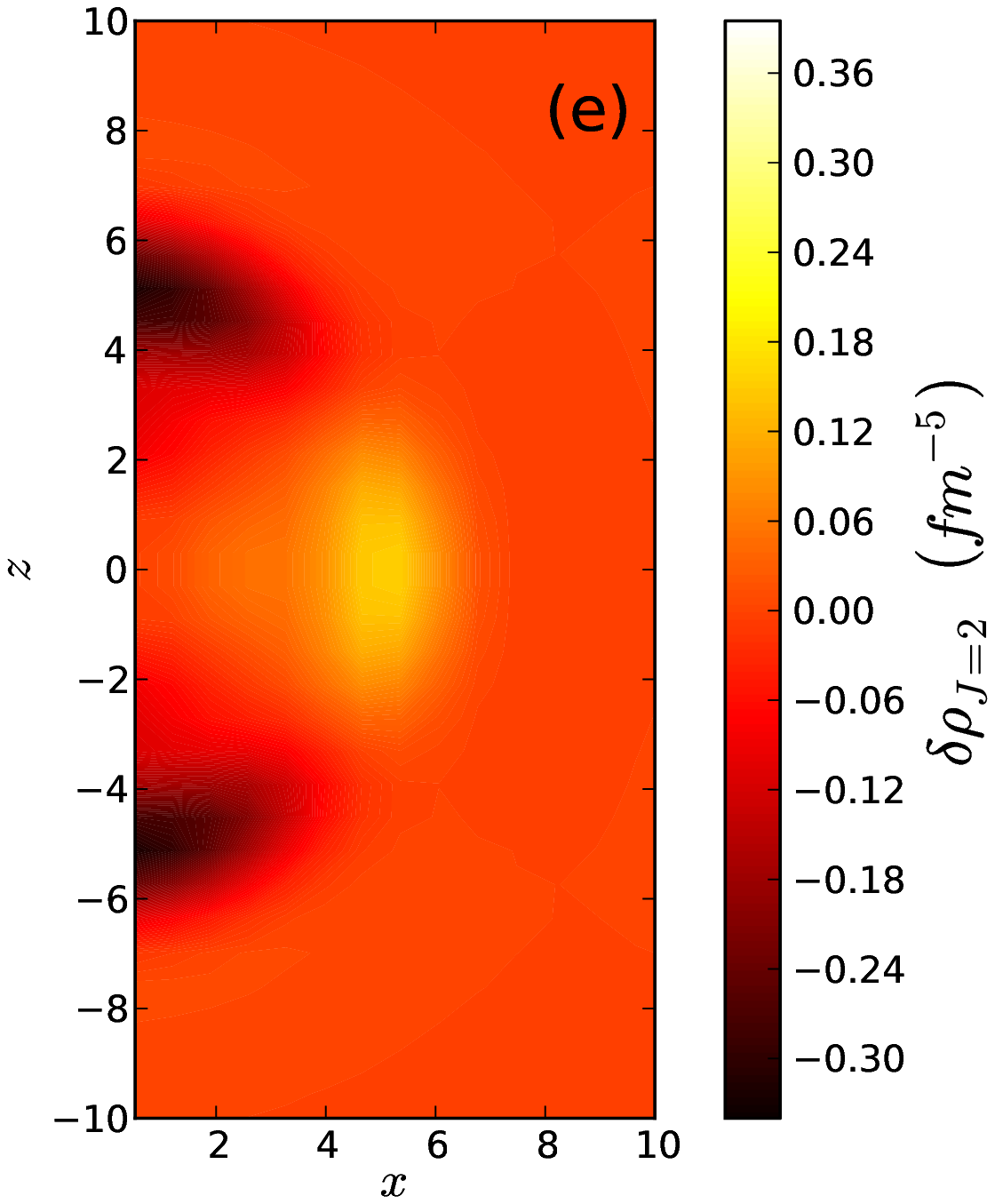} \\
\includegraphics[trim=7cm 0cm 1.5cm 1cm, clip=true, width=5.5cm]{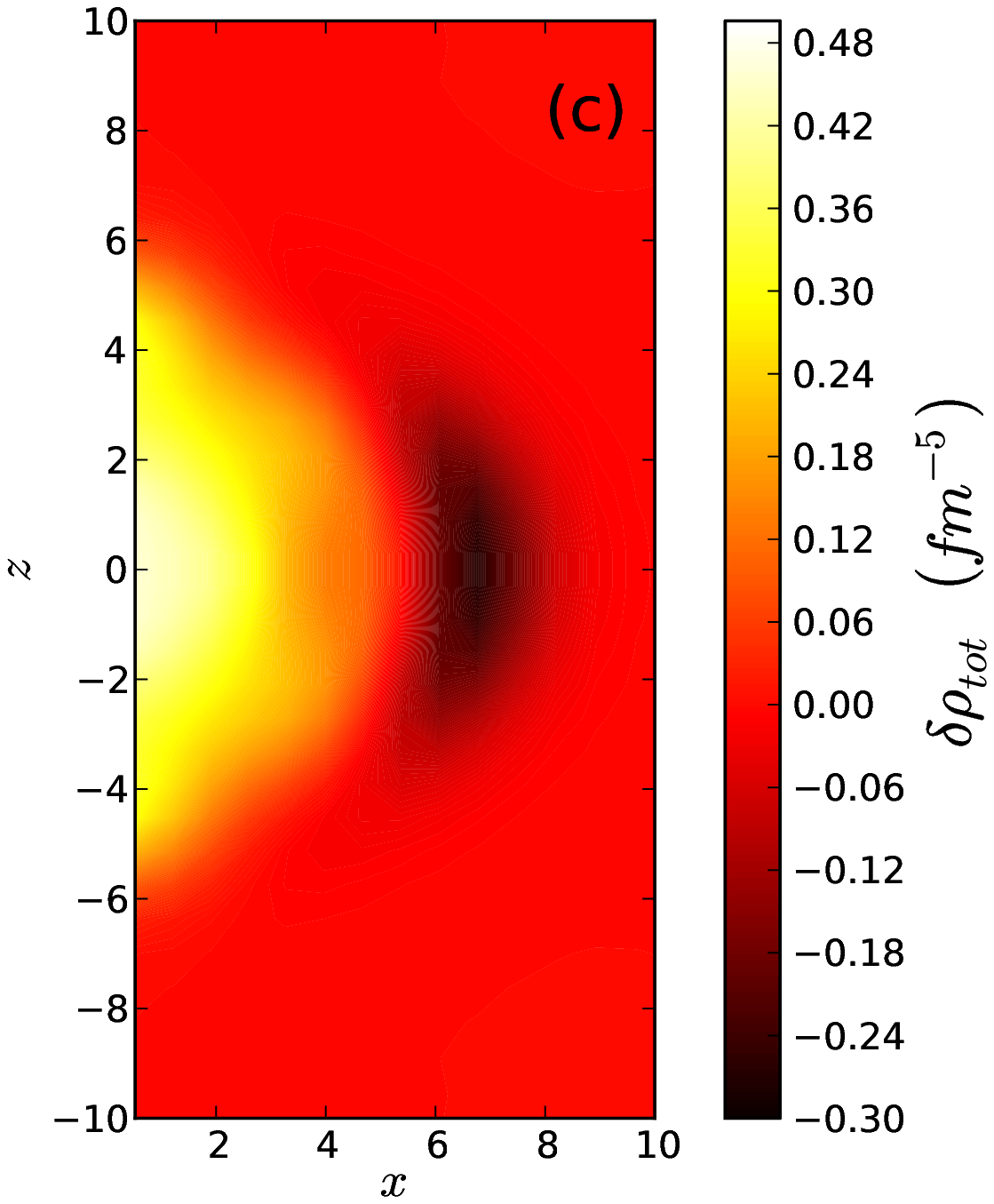} 
& \includegraphics[trim=7cm 0cm 1.5cm 1cm, clip=true, width=5.5cm]{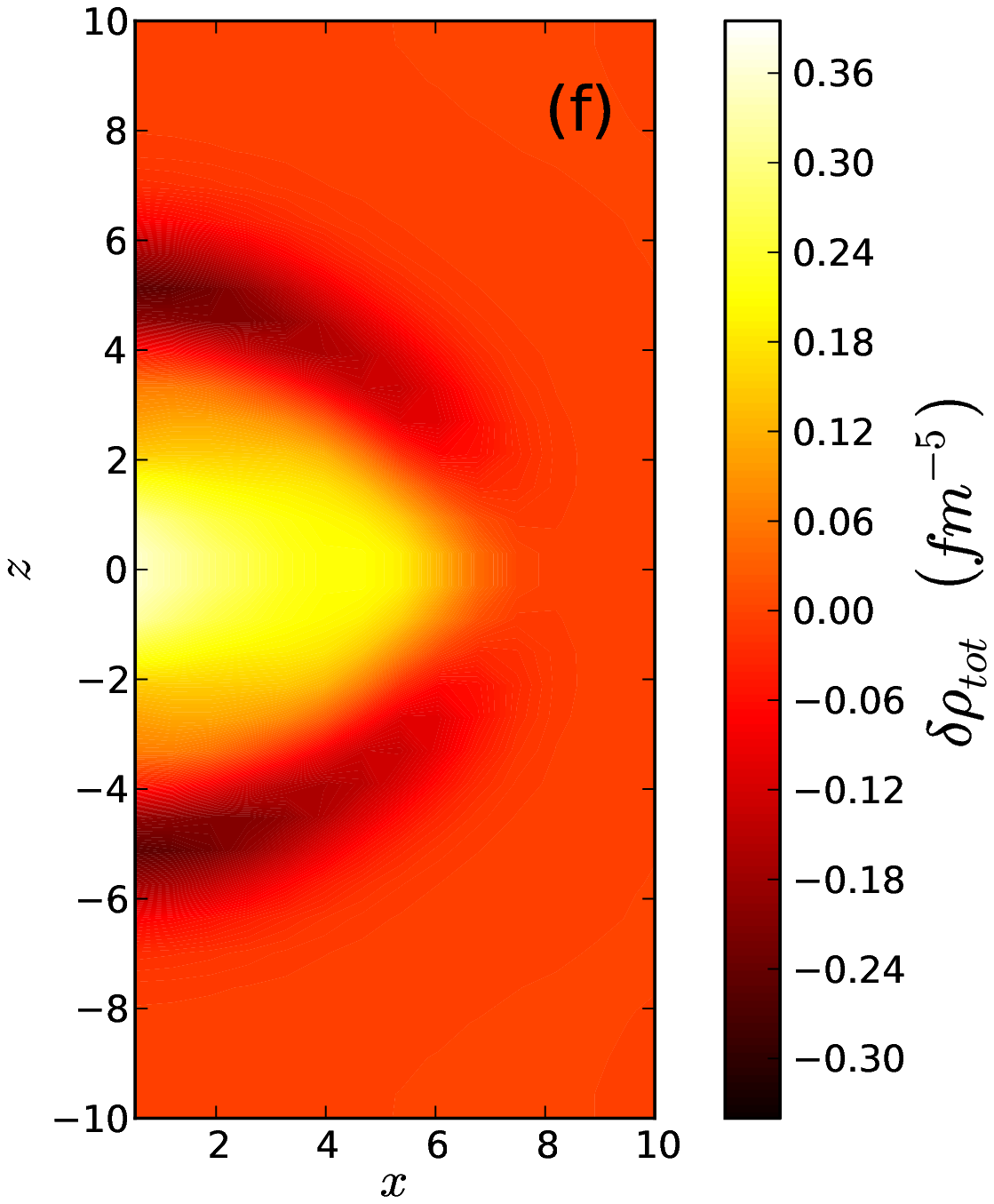}
\end{tabular}
\caption{\label{fig:sm-td-obl}(Color online) Same as in the caption to Fig.~\ref{fig:sm-td-obl} 
but for the oblate configuration with the constraint deformation $\beta=-0.3$ in $^{144}$Sm.}
\end{figure}

\begin{figure}
\centering
\includegraphics[width=\textwidth]{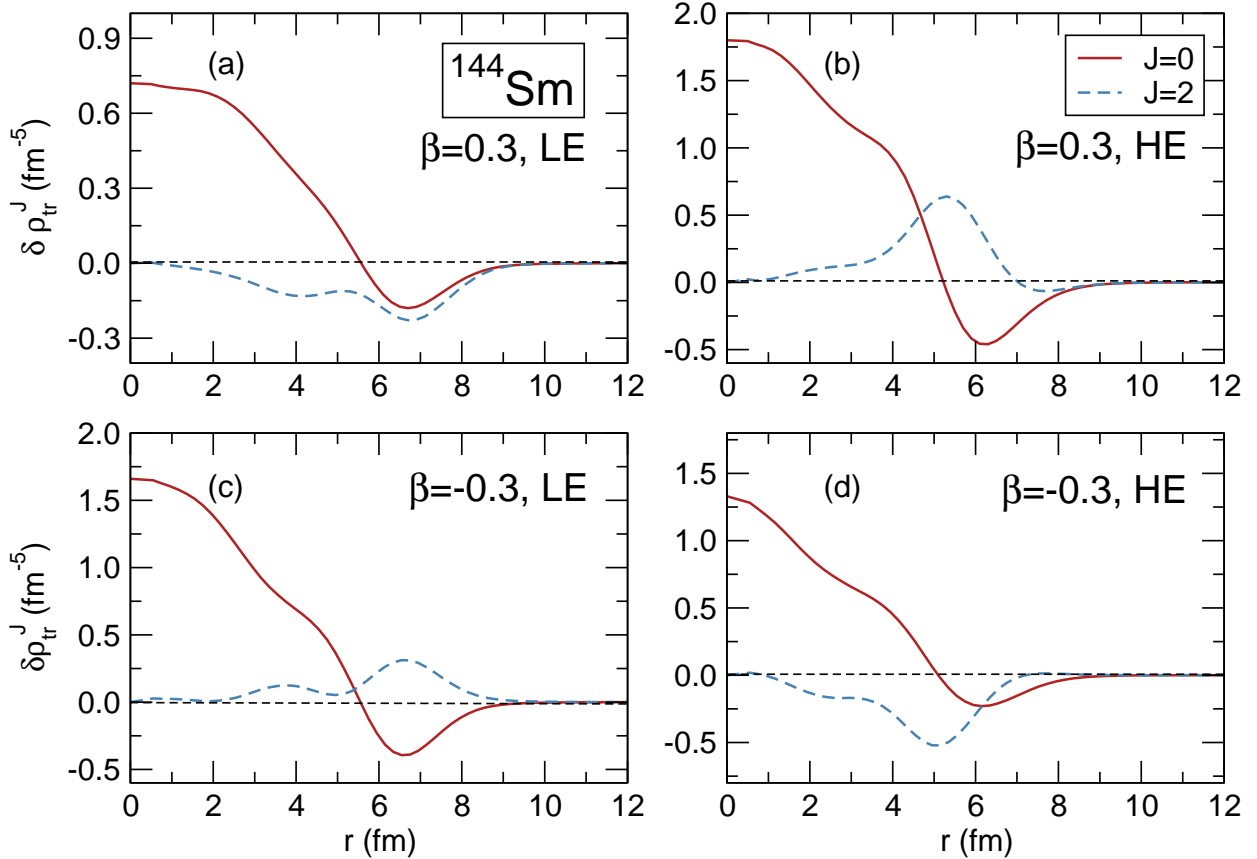} 
\caption{\label{fig:sm-transdens_rad}(Color online) 
Radial parts of the angular-momentum-projected transition densities that 
correspond to the LE and HE ISGMR peaks in the
$^{144}$Sm isotope: the prolate configuration at 
$\beta=0.3$ (panels (a) and (b)), and the oblate configuration at $\beta=-0.3$ 
(panels (c) and (d)). }
\end{figure}

Individual modes of collective excitations can be studied qualitatively by analyzing the 
corresponding transition densities. For $K^\pi=0^+$  
the intrinsic transition densities are axially symmetric:
\begin{equation}
\delta \rho_{tr}(\mathbf{r}) = \delta \rho_{tr}(r_\perp,z).
\end{equation}
By projecting the two dimensional intrinsic transition densities $\delta \rho_{tr}(r_\perp, z)$ onto good angular momentum, one obtains the transition densities in the laboratory frame of reference.
For a particular value of the angular momentum $J\ge K$, the projected transition density reads
\begin{equation}
\delta \rho_{tr}^J(\mathbf{r}) = \delta \rho_{tr}^J(r) Y_{JK}(\Omega) \;,
\end{equation}
with the radial part of the projected transition density
\begin{equation}
\delta \rho_{tr}^J(r)  = \int{d\Omega \delta \rho_{tr}(r_\perp,z) Y_{JK}(\Omega)}\;.
\end{equation} 
Although the last equation is not exact, it yields accurate results for large deformations.
As an example we chose two axially-constrained deformed configurations of $^{144}$Sm: 
the oblate configuration at deformation $\beta=-0.3$, and the prolate configuration at $\beta=0.3$.
The $J=0$ and $J=2$ angular-momentum-projected transition densities, 
and the intrinsic transition densities for the LE and HE peaks of the ISGMR strength 
distributions are shown in  Figs.~\ref{fig:sm-td-pro} (prolate deformed configuration, $\beta=0.3$) 
and~\ref{fig:sm-td-obl} (oblate deformed  configuration, $\beta=-0.3$). 
The left and right columns in Figs.~\ref{fig:sm-td-pro} and~\ref{fig:sm-td-obl} correspond to the 
LE and HE modes, respectively.  Although both the HE and the LE mode represent a mixture
of monopole and quadrupole oscillations, one can observe distinct features. 
For the HE mode the interference between the $J=0$ and $J=2$ components
of the transition density is constructive at the poles and destructive in the
equatorial plane of the density ellipsoid of $^{144}$Sm.
Figure~\ref{fig:sm-transdens_rad} 
compares the radial parts of the angular momentum projected transition densities
$\delta \rho_{tr}^{J=0}(r)$ and $\delta \rho_{tr}^{J=2}(r)$
that correspond to the LE and HE peaks in the $^{144}$Sm isotope: the prolate configuration at 
$\beta=0.3$ (panels (a) and (b)), and the oblate configuration at $\beta=-0.3$ 
(panels (c) and (d)). 
The $\delta \rho_{tr}^{J=0}(r)$ component displays the characteristic 
radial dependence of the monopole (compression) transition strength with a 
single node in the surface region. Both the volume and surface contributions are more pronounced  
for the HE component at prolate deformation, whereas for the oblate deformed configuration 
the LE component dominates. This is consistent with the strength
distributions displayed in Fig.~\ref{fig:evolution-sm144}. In all cases 
$\delta \rho_{tr}^{J=2}(r)$ has a radial dependence characteristic for 
quadrupole oscillations. We also notice that for the
LE component the surface contributions of the 
$\delta \rho_{tr}^{J=0}(r)$ and $\delta \rho_{tr}^{J=2}(r)$ transition densities are in phase
when the nucleus has prolate deformation, and out of phase when the deformation is
oblate. The opposite is found for the HE energy component.


\section{\label{sec-summary} Summary and outlook}
Realistic QRPA calculations for deformed nuclei still present a considerable computational
challenge, particularly if one considers heavy nuclei. The dimension of the configuration
space increases rapidly in heavier open-shell nuclei, and thus it becomes increasingly 
difficult to compute and store huge QRPA matrices. Although several relativistic
QRPA studies have been performed for axially deformed nuclei, computationally 
this task is simply to complex for systematic large scale calculations.
One possible solution is to employ the finite amplitude method in the solution 
of the corresponding linear response problem.
In this work we have implemented a recently proposed efficient method for the
iterative solution of the FAM -- QRPA equations in the framework of  
relativistic energy density functionals. 
Several numerical tests have been performed to verify the stability of the 
FAM-RQRPA iterative solution and its consistency with the solution of the matrix
RQRPA equations. As an illustrative example,
the FAM-RQRPA model has been applied to an analysis of the splitting of the
giant monopole resonance in deformed nuclei. In particular, we have
investigated the evolution of the $K^\pi=0^+$ strength in the chain of samarium isotopes,
from the proton-rich $^{132}$Sm, to systems with considerable neutron excess close to 
$^{160}$Sm. To study the splitting of the monopole strength in more detail, we have
performed a deformation-constrained FAM-RQRPA calculation for the nucleus  
$^{144}$Sm. A significant mixing of the monopole and quadrupole modes
has been found for both the low-energy and high-energy 
components of the $K^\pi=0^+$ strength, consistent
with the standard interpretation of the splitting of monopole strength in deformed nuclei. 

The advantage of developing and employing the FAM-RQRPA formalism would, 
of course, be limited unless it is extended to higher multipoles. This
development is already in progress. Another important extension of the FAM-RQRPA
model is the one to the charge-exchange channel.
The FAM for the charge-exchange RQRPA
will enable the description and modeling of a variety of astrophysically relevant 
weak-interaction processes, in particular beta-decay, electron capture, and
neutrino reactions in deformed nuclei. Since many isotopes that are crucial for 
the process of nucleosynthesis display considerable deformation, a 
reliable modeling of elemental abundances necessitates a
microscopic and self-consistent description of underlying transitions and 
weak-interaction processes, and this can be attained using the charge-exchange
extension of the framework introduced in this work.
 

\begin{acknowledgements}
We thank T. Nakatsukasa for providing the spherical FAM-RPA code,
and M. Kortelainen for helpful discussions. 
This work was supported in part by the MZOS - project 1191005-1010, and the DFG cluster of excellence \textquotedblleft Origin and Structure of the Universe\textquotedblright\ (www.universe-cluster.de). T. N. 
and N. K. acknowledge support by the Croatian Science Foundation.
\end{acknowledgements}
\appendix
\section{\label{Sec-App1} The single-nucleon basis}
The Dirac single-nucleon spinors are expanded in the basis of eigenfunctions
of an axially symmetric harmonic oscillator in cylindrical coordinates:
\begin{equation}
\Phi_{\alpha}(r_\perp, z, \phi, s) = \phi_{n_z}(z) \phi_{n_\perp}^\Lambda(r_\perp)
\frac{1}{\sqrt{2\pi}}e^{i\Lambda \phi} \chi_{m_s} , 
\quad \alpha\equiv \{ \Lambda n_\perp n_z m_s\}\;,
\end{equation}
where
\begin{equation}
\phi_{n_z}(z) = \frac{N_{n_z}}{\sqrt{b_z}}H_{n_z}(\xi) e^{-\xi^2/2},\quad \xi=z/b_z, \quad
N_{n_z}=\frac{1}{\sqrt{\sqrt{\pi}2^{n_z}n_z!  }},
\end{equation}
with the Hermite polynomial $H_{n_z}(z)$, and for the $r_\perp$ coordinate
\begin{equation}
\phi_{n_\perp}^\Lambda (r_\perp) = \frac{N_{n_\perp}^\Lambda}{b_\perp}\sqrt{2}
\eta^{\Lambda/2} L_{n_\perp}^\Lambda(\eta) e^{-\eta/2},\quad
\eta  = \frac{r_\perp^2}{b_\perp^2}, \quad N_{n_\perp}^\Lambda = 
\sqrt{\frac{n_\perp!}{(n_\perp+\Lambda)!}},
\end{equation}
with the Laguerre polynomial $L_{n_\perp}^\Lambda(\eta)$.
The time-reversed state reads
\begin{equation}
\Phi_{\bar{\alpha}}(r_\perp, z, \phi, s) = \hat{T} \Phi_{\alpha}(r_\perp, z, \phi, s)
= \phi_{n_z}(z) \phi_{n_\perp}^\Lambda(r_\perp)
\frac{1}{\sqrt{2\pi}}e^{-i\Lambda \phi} (-1)^{1/2-m_s} \chi_{-m_s} \;.
\end{equation}
\section{\label{Sec-App2} Matrix elements of the monopole operator }
The matrix elements of the monopole operator can be calculated analytically in the
harmonic oscillator basis. After separating the spin and angular parts of the matrix element,
the following expression is obtained:
\begin{equation}
f_{\alpha \alpha^\prime} = 
\delta_{m_s,m_s^\prime }\delta_{\Lambda \Lambda^\prime } \int_{-\infty}^{\infty}{dz\int_0^\infty{dr_\perp r_\perp \phi^\Lambda_{ n_\perp }(r_\perp)\phi_{ n_z }(z) (r_\perp^2+z^2)
 \phi^{\Lambda^\prime}_{ n_\perp^\prime }(r_\perp)\phi_{ n_z^\prime }(z)  } }  .
\end{equation}
Using the orthogonality of the eigenfunctions in the $z$ and $r_\perp$ coordinates, the matrix 
element can be written as 
\begin{equation}
f_{\alpha \alpha^\prime} = \delta_{m_s,m_s^\prime }\delta_{\Lambda \Lambda^\prime }
\times \left[\delta_{n_zn_z^\prime} \int_0^\infty{dr_\perp  r_\perp^3 
\phi^\Lambda_{ n_\perp }(r_\perp)  \phi^{\Lambda^\prime}_{ n_\perp^\prime }(r_\perp)}
+ \delta_{n_\perp n_\perp^\prime} \int_{-\infty}^\infty{dz   
z^2\phi_{ n_z }(z)  \phi_{ n_z^\prime }(z)} \right] \;.
\end{equation}
For the integral over $z$ direction one uses the recursive relation
\begin{equation}
\xi^2H_{n_z^\prime}(\xi) = \frac{1}{4}H_{n_z^\prime+2}(\xi)
+\frac{1}{2}(2n_z^\prime+1) H_{n_z^\prime}(\xi) + n_z^\prime (n_z^\prime-1)H_{n_z^\prime-2}(\xi)\;,
\end{equation}
which can be expressed in terms of harmonic oscillator eigenfunctions
\begin{equation}
\xi^2 \phi_{n_z^\prime}(\xi) 
=  \frac{1}{2} \sqrt{(n_z^\prime+2)(n_z^\prime+1)}\phi_{n_z^\prime+2}(\xi)
+\frac{1}{2}(2 n_z^\prime +1) \phi_{n_z^\prime}(\xi)
+\frac{1}{2}\sqrt{n_z^\prime(n_z^\prime-1)}\phi_{n_z^\prime-2}(\xi).
\end{equation}
This yields
\begin{equation}
I_z= \int_{-\infty}^\infty{ \phi_{ n_z }(z)z^2  \phi_{ n_z^\prime }(z)dz }
=\left\{ \begin{array}{ll} \frac{1}{2}\sqrt{n_z^\prime(n_z^\prime-1)} b_z^2, & n_z=n_z^\prime-2 \\
   \left( n_z + \frac{1}{2} \right) b_z^2, & n_z=n_z^\prime \\
    \frac{1}{2}\sqrt{n_z(n_z-1)} b_z^2, & n_z=n_z^\prime+2 \\
    0, & \textrm{otherwise} \end{array} \right. .
\end{equation}
To calculate the integral in $r_\perp$ the recursive relation
\begin{equation}
\eta L_{n^\prime}^\Lambda(\eta) = (2n^\prime + \Lambda +1)  L_{n^\prime}^\Lambda(\eta)
-(n^\prime+1)L_{n^\prime+1}^\Lambda(\eta) 
- (n^\prime+\Lambda)L_{n^\prime-1}^\Lambda(\eta) ,
\end{equation}
can expressed in the form
\begin{align}
\eta \phi_{n_\perp^\prime}^\Lambda(r_\perp)
=(2n_\perp^\prime+\Lambda+1) \phi_{n_\perp^\prime}^\Lambda(r_\perp)
&-\sqrt{n_\perp^\prime(n_\perp^\prime+\Lambda)}  \phi_{n_\perp^\prime-1}^\Lambda(r_\perp)
\\ \nonumber
&-\sqrt{(n_\perp^\prime+1)(n_\perp^\prime+\Lambda+1)}  \phi_{n_\perp^\prime+1}^\Lambda(r_\perp)\;.
\end{align}
Using this relation one obtains
\begin{equation}
I_\perp = \int_0^\infty{ \phi_{n_\perp}^\Lambda(\eta) \eta \phi_{n_\perp^\prime}^\Lambda(\eta)d\eta  }
= \left\{ \begin{array}{ll} 
(2n_\perp + \Lambda + 1)b_\perp^2 , & n_\perp^\prime = n_\perp \\
-\sqrt{n_\perp^\prime(n_\perp^\prime+\Lambda)}b_\perp^2, & n_\perp^\prime = n_\perp+1 \\
-\sqrt{n_\perp(n_\perp+\Lambda)}b_\perp^2 , & n_\perp^\prime = n_\perp-1 \\
0 & \textrm{otherwise} \end{array}  \right. 
\end{equation}
The monopole operator does not mix states from different $K^\pi$ blocks, and the matrix elements 
are real and symmetric. 
\section{\label{Sec-App3} Time-odd terms}
The time-odd current reads
\begin{equation}
\mathbf{j}(\mathbf{r}) = \sum_{\alpha \tilde{\alpha}}{\left[
\rho_{\alpha \tilde{\alpha}}\Phi_{\alpha}^\dagger\boldsymbol{\sigma}\Phi_{\tilde{\alpha}}
+\rho_{\tilde{\alpha} \alpha}\Phi_{\tilde{\alpha}}^\dagger\boldsymbol{\sigma}\Phi_{\alpha}
\right] },
\end{equation}
where $\alpha$ ($\tilde{\alpha}$) denotes the harmonic oscillator quantum numbers for the 
large (small) component of the single-nucleon Dirac spinor. 
The $\boldsymbol{\sigma}$ matrix can be expressed in cylindrical coordinates
\begin{equation}
\boldsymbol{\sigma} = e^{-i\phi} \sigma_+ \boldsymbol{e_\perp}
+e^{i\phi} \sigma_- \boldsymbol{e_\perp}
- i e^{-i\phi} \sigma_+ \boldsymbol{e_\phi} 
+ i e^{i\phi} \sigma_- \boldsymbol{e_\phi}
+\sigma_z \boldsymbol{e_z} \;.
\end{equation}
The following expressions can easily be evaluated
\begin{align}
\Phi_\alpha^\dagger \sigma_+ e^{-i\phi}\Phi_\beta &=
\frac{1}{2\pi} \delta_{m_s^\alpha,1/2}\delta_{m_s^\beta,-1/2}
\phi_{n_z^\alpha}(z) \phi_{n_z^\beta}(z) \phi^{\Lambda_\alpha}_{n_\perp^\alpha}(r_\perp)
 \phi^{\Lambda_\beta}_{n_\perp^\beta}(r_\perp) e^{i(\Lambda_\beta-\Lambda_\alpha-1)\phi},\\
 \Phi_\alpha^\dagger \sigma_- e^{i\phi}\Phi_\beta &=
 \frac{1}{2\pi} \delta_{m_s^\alpha,-1/2}\delta_{m_s^\beta,1/2}
\phi_{n_z^\alpha}(z) \phi_{n_z^\beta}(z) \phi^{\Lambda_\alpha}_{n_\perp^\alpha}(r_\perp)
 \phi^{\Lambda_\beta}_{n_\perp^\beta}(r_\perp) e^{i(\Lambda_\beta-\Lambda_\alpha+1)\phi}.
\end{align}
Next, we use the condition for monopole excitations $\Omega_\alpha=\Omega_\beta$, that is,
$\Lambda_\alpha+m_s^\alpha=\Lambda_\beta+m_s^\beta$,
\begin{align}
\Phi_\alpha^\dagger \left( \sigma_+ e^{-i\phi} + \sigma_- e^{i\phi} \right) \Phi_\beta &=
\frac{1}{2\pi} \delta_{m_s^\alpha,-m_s^\beta}
\phi_{n_z^\alpha}(z) \phi_{n_z^\beta}(z) \phi^{\Lambda_\alpha}_{n_\perp^\alpha}(r_\perp)
 \phi^{\Lambda_\beta}_{n_\perp^\beta}(r_\perp), \\
 \Phi_\alpha^\dagger \left(  \sigma_- e^{i\phi}-\sigma_+ e^{-i\phi}  \right) \Phi_\beta &=
\frac{1}{2\pi}(-1)^{1/2-m_s^\beta} \delta_{m_s^\alpha,-m_s^\beta}
\phi_{n_z^\alpha}(z) \phi_{n_z^\beta}(z) \phi^{\Lambda_\alpha}_{n_\perp^\alpha}(r_\perp)
 \phi^{\Lambda_\beta}_{n_\perp^\beta}(r_\perp) .
\end{align}
and, finally, calculate the contribution from the $z$ component
\begin{equation}
\Phi_\alpha^\dagger \sigma_z \Phi_\beta = 
\frac{1}{2\pi}(-1)^{1/2-m_s^\beta} \delta_{m_s^\alpha,m_s^\beta}
\phi_{n_z^\alpha}(z) \phi_{n_z^\beta}(z) \phi^{\Lambda_\alpha}_{n_\perp^\alpha}(r_\perp)
 \phi^{\Lambda_\beta}_{n_\perp^\beta}(r_\perp) \;.
\end{equation}
The following relations are valid
\begin{align}
\Phi^\dagger_\alpha \left( \sigma_+ e^{-i\phi} + \sigma_- e^{i\phi} \right)
\Phi_\beta &= \Phi^\dagger_\beta \left( \sigma_+ e^{-i\phi} + \sigma_- e^{i\phi} \right)
\Phi_\alpha, \\
\Phi^\dagger_\alpha \left( \sigma_- e^{i\phi} - \sigma_+ e^{-i\phi} \right)
\Phi_\beta &= -\Phi^\dagger_\beta \left( \sigma_- e^{i\phi} - \sigma_+ e^{-i\phi} \right)
\Phi_\alpha,\\
\Phi^\dagger_\alpha \sigma_z\Phi_\beta &= \Phi^\dagger_\beta \sigma_z\Phi_\alpha \;,
\end{align}
and also
\begin{align}
\Phi^\dagger_{\bar{\alpha}} \left( \sigma_+ e^{-i\phi} + \sigma_- e^{i\phi} \right)
\Phi_{\bar{\beta}} &=- \Phi^\dagger_\alpha \left( \sigma_+ e^{-i\phi} + \sigma_- e^{i\phi} \right)
\Phi_\beta, \\
\Phi^\dagger_{\bar{\alpha}} \left( \sigma_- e^{i\phi} - \sigma_+ e^{-i\phi} \right)
\Phi_{\bar{\beta}} &= \Phi^\dagger_\alpha \left( \sigma_- e^{i\phi} - \sigma_+ e^{-i\phi} \right)
\Phi_\beta,\\
\Phi^\dagger_{\bar{\alpha}} \sigma_z\Phi_{\bar{\beta}} &= 
-\Phi^\dagger_\alpha \sigma_z\Phi_\beta.
\end{align}
The corresponding elements of the Hamiltonian matrix read
\begin{equation}
\langle \alpha | \boldsymbol{\sigma} \cdot \mathbf{V} | \beta \rangle
= \langle \alpha | (e^{-i\phi}\sigma_+ + e^{i\phi}\sigma_- ) V_\perp
 + i  (e^{i\phi}\sigma_- - e^{-i\phi}\sigma_+) V_\phi + \sigma_z V_z | \beta\rangle.
\end{equation}
\section{\label{Sec-App4} FAM equations for time-reversal symmetry}
We consider systems with time-reversal, reflection and axial  symmetries. 
The single-quasiparticle states can be ordered so that we first list states 
with $\Omega > 0$, and then states with 
$\Omega <0 $. The HFB matrices $U$ and $V$ read 
\begin{equation}
U = \left( \begin{array}{cc} u & 0 \\ 0 & u^* \end{array}\right), \quad
V = \left( \begin{array}{cc} 0 & -v^* \\ v & 0 \end{array} \right) .
\end{equation}
This generates a density matrix and pairing tensor with block-diagonal structure
\begin{align}
\rho &= V^* V^T = \left( \begin{array}{cc} 0 & v^* \\ -v & 0\end{array} \right)^* 
\left( \begin{array}{cc} 0 & v^* \\ -v & 0\end{array} \right)^T =
\left( \begin{array}{cc} vv^\dagger & 0 \\ 0 & v^* v^T   \end{array} \right)
= \left( \begin{array}{cc} \rho_1 & 0 \\ 0 & \rho_2 \end{array} \right),\\
\kappa &= V^* U^T =  \left( \begin{array}{cc} 0 & v^* \\ -v & 0\end{array} \right)^* 
\left( \begin{array}{cc} u & 0 \\ 0 & u^* \end{array} \right)^T
= \left( \begin{array}{cc}  0 & v u^\dagger \\ -v^* u^T & 0\end{array} \right)
=  \left( \begin{array}{cc} 0 & \kappa_2 \\ \kappa_1 & 0 \end{array} \right).
\end{align}
The FAM amplitudes $X$ and $Y$ are antisymmetric matrices
\begin{equation}
X = \left(  \begin{array}{cc} 0 & x \\ -x^T & 0 \end{array} \right), \quad
Y = \left( \begin{array}{cc} 0 & y \\ -y^T & 0 \end{array} \right) ,
\end{equation}
where $x$ and $y$ are symmetric complex matrices.
The explicit expressions for the density matrix and pairing tensor read
\begin{equation}
\begin{array}{ll}
\rho_1 = \left( v - \eta u x \right) \left( v-\eta u y^* \right)^\dagger , \quad
&\rho_2 = \left( v - \eta u x^\dagger \right)^* \left( v-\eta u y^T \right)^T , \\
\kappa_1 = \left( v-\eta u x^\dagger \right)^* \left( u+\eta v y^T \right)^T,\quad
&\kappa_2 =- \left( v-\eta u x \right) \left( u+\eta v y^*\right)^\dagger,\\
\bar{\kappa}_1 = \left( v - \eta u y^T \right)^*\left( u+\eta vx^\dagger \right)^T , \quad
&\bar{\kappa}_2 = - \left( v - \eta u y^* \right) \left( u+\eta v x \right)^\dagger .
\end{array}
\end{equation}
It should be noted that since the $x$ and $y$ matrices are complex, the relations $\rho_2 = \rho_1^*$
and $\kappa_i^\dagger = \kappa_i$ are no longer fulfilled.
The matrices  $H^{20}(\omega)$ and $H^{02}(\omega)$ read
\begin{equation}
\delta H^{20}(\omega) = \left( \begin{array}{cc}
0 &  \delta h^{20} \\ -\left[ \delta h^{20}\right]^T & 0
\end{array} \right), \quad
\delta H^{02}(\omega) = \left( \begin{array}{cc}
0 &  \delta h^{02} \\ -\left[ \delta h^{02}\right]^T & 0
\end{array} \right)
\end{equation}
with
\begin{align}
\delta h^{20}(\omega) &= -u^\dagger \delta h_1 v + u^\dagger \delta \Delta_2 u 
+ v^\dagger \delta \bar{\Delta}_1 v - v^\dagger \delta h_2^T u ,\\
\delta h^{02}(\omega) &= v^T \delta h_2 u^* - u^T \delta \bar{\Delta}^*_2 u^*
    -v^T \delta \Delta_1 v^* + u^T \delta h_1^T v^*.
\end{align}
The matrices $F^{20}$ and $F^{02}$ of the external operator are decomposed in
an analogous way.
Time-reversal symmetry reduces by half the dimension of the equations of motion 
\begin{align}
(E_\mu+E_\nu -\omega) x_{\mu \nu} + \delta h^{20}_{\mu \nu} + f^{20}_{\mu \nu} &= 0,\\
 (E_\mu+E_\nu +\omega)y_{\mu \nu} + \delta h^{02}_{\mu \nu} + f^{02}_{\mu \nu} &= 0 .
\end{align}

%


\end{document}